\documentclass{article}

\usepackage{amsfonts} 
\usepackage{graphics}
\usepackage{amsmath}
\usepackage{amscd}
\usepackage{amssymb}
\usepackage{rlepsf}

\usepackage{color}

\usepackage{euscript}

\newtheorem{Theorem}{Theorem}[section]

\newtheorem{Definition}[Theorem]{Definition}

\newtheorem{Lemma}[Theorem]{Lemma}

\newtheorem{Fundamental Theorem}{Fundamental Theorem}

\newenvironment{Proof}[1][Proof]{\textbf{#1.} }{\ \rule{0.5em}{0.5em}}

\def \V { V_\TV}
\def \v {v_\TV}
\def \C {\mathbb{C}}
\def \CH {{\mathrm{CH}}}
\def \D {\Delta}
\def \d {\partial}
\def \e {\epsilon}
\def \g {\gamma}
\def \Gc {{\cal G}}
\def \i {{\cal I}}
\def \l{\lambda}
\def \G {\Gamma}
\def \O {\mathcal{O}}

\def \R {\mathbb{R}}
\def \S {\Sigma}
\def \s {\scriptstyle}
\def \sq {\square}
\def \n {\eta}

\def \T {\mathrm{T}}
\def \Tc {{\cal T}}
\def \TV  {{\mathrm{TV}}}
\def \W {\Omega}
\def \WRT {{\mathrm{WRT}}}
\def \Y {\rm Y}
\def \N {\mathbb{N}}
\def \Z {\mathbb{Z}}
\begin{document}

\title{Spin Foam Perturbation Theory for Three-Dimensional Quantum Gravity}

\author{Jo\~{a}o  Faria Martins \\ \footnotesize\it  {Centro de Matem\'{a}tica da Universidade do Porto,}\\ {\footnotesize\it Departamentos de
Matem\'{a}tica}\\{\footnotesize\it
Rua do Campo Alegre, 687, 4169-007 Porto, Portugal}\\ {\footnotesize\it jnmartins@fc.up.pt} \\ \\Aleksandar Mikovi\'c \footnote{Member of the Mathematical Physics Group, University of Lisbon}\\
{\footnotesize \it{Departamento de Matem\'atica}}\\ {\footnotesize\it {Universidade Lus\'{o}fona de Humanidades e Tecnologia,}}\\ {\footnotesize \it{Av do Campo Grande, 376, 1749-024, Lisboa, Portugal}}\\ {\footnotesize \it {amikovic@ulusofona.pt}}}

\maketitle
\begin{abstract}
{We {formulate} the spin foam perturbation theory for three-dimensional Euclidean Quantum Gravity {with a cosmological constant}. We analyse the perturbative expansion of the partition function in the dilute-gas limit and we argue that the Baez conjecture {stating that the number of possible distinct topological classes of perturbative configurations is finite for the set of all  triangulations of a manifold,} is not true. However, the conjecture {is true} for a special class of triangulations which are based on subdivisions of certain 3-manifold cubulations. In this case we calculate the partition function {and show that the dilute-gas correction vanishes for the simplest choice of the volume operator. By slightly modifying the dilute-gas limit, we obtain a nonvanishing correction which is related to the second order perturbative correction. By assuming that the dilute-gas limit coupling constant is a function of the cosmological constant, we obtain a value for the partition function which is independent of the choice of the volume operator.}}
\end{abstract}
\newpage
\tableofcontents

\section{Introduction}

{Spin foam state sum models can be understood as the path integrals for  BF topological field theories \cite{Ba}.} Since General Relativity in 3 and 4 dimensions can be represented as a perturbed BF theory, see \cite{FK,M2}, then, in order to find the corresponding Quantum Gravity theory, one would need a spin foam perturbation theory. Baez has analysed the spin foam perturbation theory from a general point of view in \cite{Ba2}, and he was able to show that, under certain reasonable assumptions, the perturbed spin foam state sum $Z$ can be calculated in the dilute gas limit. In this limit, {the number $N$ of {tetrahedra}} of a manifold triangulation $\Delta$ tends to infinity and $\lambda$, the perturbation theory parameter, tends to zero, {in a way such that the effective coupling constant $g=\lambda N$  is finite.} By assuming that the number of topologically inequivalent classes of perturbed configurations at a given order of perturbation theory is limited when $N \to\infty$, Baez showed that the perturbation series 
\begin{equation} 
Z(M,\Delta) = Z_0(M) + \lambda Z_1(M,\Delta) + \lambda^2 Z_2(M,\Delta) + \cdots =\sum_{n=0}^\infty \lambda^n Z_n (M,\Delta) \,,\label{ps} 
\end{equation}  
where $M$ is the manifold, is dominated by the contributions from the dilute configurations in the dilute gas limit. The dilute configurations of order $n$ are the configurations where $n$ non-intersecting simplices carry a single perturbation insertion. Let $\bar Z_N = \sum_{n=0}^N \lambda^n Z_n (M,\Delta)$, then 
\begin{equation}
\lim_{N\to\infty}\bar Z_N(M,\Delta) = e^{gz_1} Z_0 (M)\,, \label{dglf}
\end{equation}
where 
$$z_1 =\lim_{N\to\infty} {Z_1(M,\D) \over N Z_0(M)}$$ 
does not depend on the chosen manifold $M$.

In this paper, we are going to study in detail the Baez approach on the example of three-dimensional (3d) Euclidean Quantum Gravity {with a cosmological constant}. In this case it is possible to construct explicitly the perturbative corrections, {and we will show how to do it}. Therefore one can check all the assumptions and the results from the general approach. We will show that the conjecture that there are {only} finitely many topological classes of perturbative configurations at a given order of the perturbation theory is not true. However, if the triangulations are restricted to those corresponding to special subdivisions of certain acceptable manifold cubulations, then the number of  these topological classes is finite. In this case we show that the formula (\ref{dglf}) is still valid and we calculate $z_1$. A surprising feature of 3d gravity is that $z_1 =0$ and consequently the dillute gas limit has to be modified in order to obtain a nonzero contribution. We will show that for ${g=\lambda^2 N}$
\begin{equation}
\lim_{N\to\infty}\bar Z_N(M,\Delta) = e^{gz_2} Z_0 (M)\,, \label{mdgl}
\end{equation}
where $${z_2 =\lim_{N\to\infty} { Z_2 (M,\Delta) \over N Z_0 (M)} \,.}$$
{The value of $z_2$}, {which we conjecture to be non-zero}, is independent of { the triangulation $\D$ and} the manifold $M$. {{Recall that $N$ denotes the number of tetrahedra of the triangulation $\D$ of $M$.}

{In order to construct the perturbative corrections $Z_n$, we will use the path-integral expression for the partition function of  Euclidean 3d gravity with a cosmological constant $\lambda$}
\begin{equation}
Z(M,\Lambda)= \int {\cal D}A {\cal D}B \exp\left(i\int_M  {\rm Tr}\,(B\wedge F) + \Lambda\, \epsilon_{abc}B^a\wedge B^b \wedge B^c \right) \,,\label{3dpi}
\end{equation}
where $A$ is an ${\rm SU}(2)$ principal bundle connection, $F$ is the corresponding curvature 2-form, $B$ is a one-form taking values in the ${{\rm SU}}(2)$ Lie algebra {and $\epsilon_{abc}$ are the structure constants}. This path integral can be defined as a finite spin foam state sum when $\Lambda = 4\pi^2 / r^2$, $r\in \N$, {see \cite{FK}}, and in this case it is given by the Turaev-Viro (TV) state sum \cite{TV}. However, if $\Lambda \ne 4\pi^2 / r^2$ then it is not obvious how to define $Z$. A natural approach is to use the generating functional technique \cite{FK}, and in \cite{HS} the first order perturbation theory spin foam amplitudes were studied for the Ponzano-Regge (PR) model \cite{PR}. However, the problem with the PR model is that it is not finite, so that the state sums $Z_n$ in (\ref{ps}) are not well defined. Since the TV model can be considered as a quantum group regularisation of the PR model, we are going to use the TV model to define the perturbation series (\ref{ps}). Physically this means expanding the path integral (\ref{3dpi}) by using $\lambda=\Lambda -4\pi^2 / r^2$ as the perturbation theory parameter instead of $\lambda=\Lambda$. 

The TV model perturbation series can be constructed by using the PR model perturbation series and then replacing all the weights in the PR amplitudes with the corresponding quantum group spin network evaluations. The calculation of the corresponding state sums is substantially simplified if the Chain-Mail technique is used, see \cite{R1,BGM,FMM}. In section 2 we review the PR perturbation theory. In section 3 we review the Chain-Mail technique, while in section 4 we define the perturbative corrections. In section 5 we discuss the dilute gas limit, while in section 6 we present our conclusions.

\section{Perturbative expansion for the PR model}\label{Pert}

Given a triangulation $\Delta$ of $M$ with $N$ tetrahedra, let us associate to each edge $\e$ of $\Delta$ a source current $J_{\e}= J^a_{\e} T_a$ which belongs to the Lie algebra ${{\mathfrak{su}}}(2)$ with a basis $\{ T_a |a=1,2,3\}$. One can then write
\begin{equation}
 Z = \exp\left(-\lambda\sum_{k=1}^N \d^3_J (\tau_k) \right)Z(J){\Big |}_{J=0}\,,\label{gff}
\end{equation}  
where $\d^3_J (\tau_k)$ is a differential operator associated with the volume of a tetrahedron $\tau_k$ and $Z(J)$ is the generating functional, given by the Ponzano-Regge state sum with the $D^{(j_{\e})}(e^{J_{\e}})$ insertions at the edges of the $6j$ spin networks, where $D^{(j)}(e^{J})$ is the matrix of the group element $e^J$ in the representation of spin $j$, see \cite{FK,HS}. The operator $\d^3_J$ can be chosen to be
$$ \d^3_J = {1\over 4}\sum_{\e, \mu, \nu}\e^{abc}{\partial\over\partial J_\e^a}{\partial\over\partial J_\mu^b}{\partial\over\partial J_\nu^c} \,,$$
where $\e^{abc}$ is a totally antisymmetric tensor and $\e,\mu,\nu$ {are  tetrahedron edges sharing} a common vertex\footnote{One can choose a more general expression for $\d^3_J$, involving all possible triples of the edges, see \cite{HS}, but in this paper we will study the simplest possible choice.}.

Since 
$$ {\partial\over \partial J_\e^a}\cdots {\partial\over \partial J_\e^{a'}}D^{(j_{\e})}(e^{J_{\e}}) {\Big |}_{J=0} = T^{(j_\e)}_{(a} \cdots T^{(j_\e)}_{a')} \,,$$
where 
$$X_{(a_1 \cdots a_p )}=\frac{1}{p!}\sum_{\sigma\in S_p}X_{a_{\sigma(1)}}\cdots X_{a_{\sigma(p)}}\,,$$
then the result of the action of $\partial^3_J$ on a tetrahedron's vertex will be given by the grasping insertion
$$ \sum_{a,b,c}\e^{abc} T^{(j)}_a T^{(k)}_b T^{(l)}_c \,,$$
where $j,k$ and $l$ are the spins of the three edges sharing the vertex.

By using
$$ \e_{abc} = C_{abc}^{111}\,,\quad \left(T^{(j)}_a \right)_{\alpha\alpha'} ={{A}_j} C_{a\alpha\alpha'}^{1jj}\,,$$ where
$C_{\alpha\beta\g}^{jkl}$ is the intertwiner tensor for ${{\rm Hom}}(V_j \otimes V_k , V_l^*)$ {($3j$ symbol)} and {${A}_j$ is a normalisation factor given by 
\begin{equation}
{A}_j^2 = {j(j+1)(2j+1) \over \theta (1,j,j)}\,,\label{aj} 
\end{equation}
one obtains 
$$ \sum_{a,b,c}\e^{abc} \left(T^{(j)}_a \right)_{\alpha\alpha'} \left(T^{(k)}_b \right)_{\beta\beta'} \left(T^{(l)}_c \right)_{\gamma\gamma'}={{A}_j {A}_k {A}_l} \sum_{a,b,c}C^{111}_{abc}C^{1jj}_{a\alpha\alpha'}C^{1kk}_{b\beta\beta'}C^{1ll}_{c\gamma\gamma'} \,.$$
This equation implies that the evaluation of a {tetrahedral} spin network with a grasping insertion is {proportional to} the evaluation of a spin network based on a tetrahedron graph with an additional trivalent vertex whose edges carry the spin one representations and connect the 3 edges carrying the spins $j,k$ and $l$, see Figure \ref{Example}. The $Z_n$ which follows from (\ref{gff}) will be then given by a sum of ${(n+N-1)!\over n!(N-1)!}$ terms where each term corresponds to the PR state sum with $n$ graspings distributed among the $N$ tetrahedra. {The weight of a tetrahedron with $m$ graspings is given by an analogous evaluation of the ${{\rm SU}}(2)$ spin network from Figure \ref{Example} with $m$ insertions.} 

In order to make all the PR state sums $Z_n$ finite, we will replace all the ${{\rm SU}}(2)$ spin networks associated with a $Z_n$ with the corresponding quantum ${{\rm SU}}(2)$ spin networks at a root of unity. In the following sections we will show how to do this.

\begin{figure}
\centerline{\relabelbox 
\epsfysize 4cm
\epsfbox{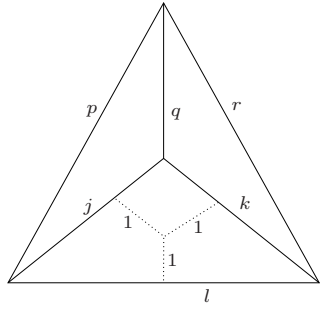}
\relabel{a}{$\s{p}$}
\relabel{b}{$\s{q}$}
\relabel{c}{$\s{r}$}
\relabel{j}{$\s{j}$}
\relabel{k}{$\s{k}$}
\relabel{l}{$\s{l}$}
\relabel{e}{$\s{1}$}
\relabel{f}{$\s{1}$}
\relabel{g}{$\s{1}$}
\endrelabelbox }\caption{\label{Example} The evaluation of a {tetrahedral} spin network with a grasping insertion. }
\end{figure}

\section{{Quantum ${\rm SU}(2)$ invariants of links and three-manifolds}}

We gather some well known facts about quantum ${\rm SU}(2)$ invariants which we will need in this paper. 

\subsection{Spin network calculus}
 
Consider an integer parameter $r \ge 3$ (fixed throughout this article), and let $q=e^{\frac{i \pi} {r}}$. Define the quantum dimensions   { $\dim_q j=(-1)^{2j}\frac{q^{2j+1} - q^{-2j-1}}{q-q^{-1}}$,} where   $i \in \{0,1/2,{\ldots}\texttt{},(r-2)/2\}$.
{If  the edges of a trivalent framed graph $\G$ embedded in $S^3$ are assigned spins $j_1,j_2,...,j_n
\in \{0,1/2,...,(r-2)/2\}$, then we can consider the value $\left
  <\G;j_1,...,j_n\right> \in \C$     obtained by}
using the quantum spin network calculus at  $q$; we will use the normalisation of \cite{KL}.
We can also  consider the case in which the edges of $\G$ are
assigned linear combination of spins, {with multilinear dependence on the
  colourings of each edge of $\G$}. A very  important linear  combination of
spins   is the ``$\W$-element'' given by: 
$$   \W= \sum_{j=0}^{\frac{r-2}{2}}   \dim_q(j) {R_j \,,}$$ 
where $R_j$ denotes the representation of spin $j$.

A sample of the  properties satisfied by  the $\W$-element appears in
\cite{R1,L,KL}. In {Figure}  \ref{lic}  we display a special case of the
Lickorish Encircling Lemma, of which we will make explicit use.

Define    $\left <\bigcirc_0;\W\right >={\n}$ and  $\left <\bigcirc_1;\W\right >=\kappa\sqrt{{\n}}$, the evaluation of the 0- and 1-framed unknots coloured with the $\W$-element. Therefore we have that {$${\n}=\sum_{j=0}^{(r-2)/2} (\dim_q j)^2= {r \over 2 \sin^2 \left ( \frac \pi r \right)}\,.$$} 
On the other hand {$\kappa= q^{\frac{-3-r^2}{2}}e^{-\frac{i \pi}{4}}$}, and 
$\left < \bigcirc_{-1}\right >=\sqrt{{\n}}\kappa^{-1}$; see \cite{R1}.

\subsection{Generalised Heegaard diagrams}

Let $M$ be a closed oriented piecewise-linear 3-manifold. Choose a handle decomposition of $M$; see \cite{RS,GS}. Let $H_-$ be the union of the 0- and 1-handles of $M$. Let also  $H_+$ be the union of the 2- and 3-handles of $M$. Both $H_-$ and $H_+$ have natural orientations induced by the orientation of $M$. There exist two non-intersecting naturally defined  framed links $m$ and $\e$ in $H_-$; see \cite{R1}. The second one is given by the attaching regions of the 2-handles of $M$ in $\d H_-=\d H_+$, pushed inside $H_-$, slightly. On the other hand, {$m$  is given by  the belt-spheres of the 1-handles of $M$, living in $\d H_-$.}  The sets of curves $m$ and $\e$  in $H_-$ have natural framings, parallel to $\d H_-$. The triple $(H_-,m,\e)$ will be called a generalised Heegaard diagram of the oriented closed 3-manifold  $M$. 

\subsection{The Chain-Mail Invariant}
We now recall the definition of J. Robert's Chain-Mail invariant of closed {oriented} $3$-manifolds. This construction will play a fundamental role in this article.
Let  $M$ be a {connected} 3-dimensional closed oriented piecewise linear  manifold. Consider a generalised Heegaard diagram $(H_-,m,\e)$, associated with a handle decomposition of $M$. Give $H_-$ the orientation induced by the orientation of $M$.  Let $\Phi\colon    H_- \to S^3$ be an orientation preserving embedding. Then the image of the links $m$ and $\e$ under $\Phi$ defines a link $\CH(H_-,m,\e,\Phi)$ in $S^3$, called the ``Chain-Mail Link''. 
   {J. Roberts} proved  that the evaluation 
$\left < \CH(H_-,m,\e,\Phi); \W \right>$
of the Chain-Mail Link coloured with the $\W$-element  is independent of the orientation preserving embedding $\Phi\colon   H_- \to S^3$; see \cite{R1}, {Proposition $3.3$. }

The Chain-Mail Invariant of $M$     is defined  as:
$$Z_\CH(M)={\n}^{-n_0-n_2}\left < \CH(H_-,m,\e,\Phi);\W \right>,$$
 where $n_i$ is the number of $i$-handles of $M$.  It is proved in \cite{R1} that this    {Chain-Mail Invariant}  is independent of the chosen handle decomposition of $M$ and that it  coincides with the Turaev-Viro Invariant $Z_\TV(M)$ of $M$; see \cite{TV}. 

\subsection{The Turaev-Viro Invariant}\label{TViro}

Let $M$ be  3-dimensional closed {connected} oriented piecewise linear manifold. Consider a piecewise linear triangulation of $M$. We can consider a  handle decomposition of $M$ where each $i$-simplex of $M$ generates a $(3-i)$-handle of $M$; see for example \cite{R1}. Applying the Chain-Mail {construction} to this handle decomposition, yields the following combinatorial  picture for the  calculation of the Chain-Mail Invariant $Z_{\CH}(M)$, which, in this form, is called  the Turaev-Viro Invariant $Z_\TV$.

A colouring of $M$ is an assignment of a spin    {$j \in \{0,1/2,...,(r-2)/2\}$} to each edge of $M$. Each colouring of a simplex $s$ gives rise to a weight $W(s)\in \C$, in the way shown in {Figure} \ref{Weight}.
\begin{figure}
\centerline{\relabelbox 
\epsfysize 10cm
\epsfbox{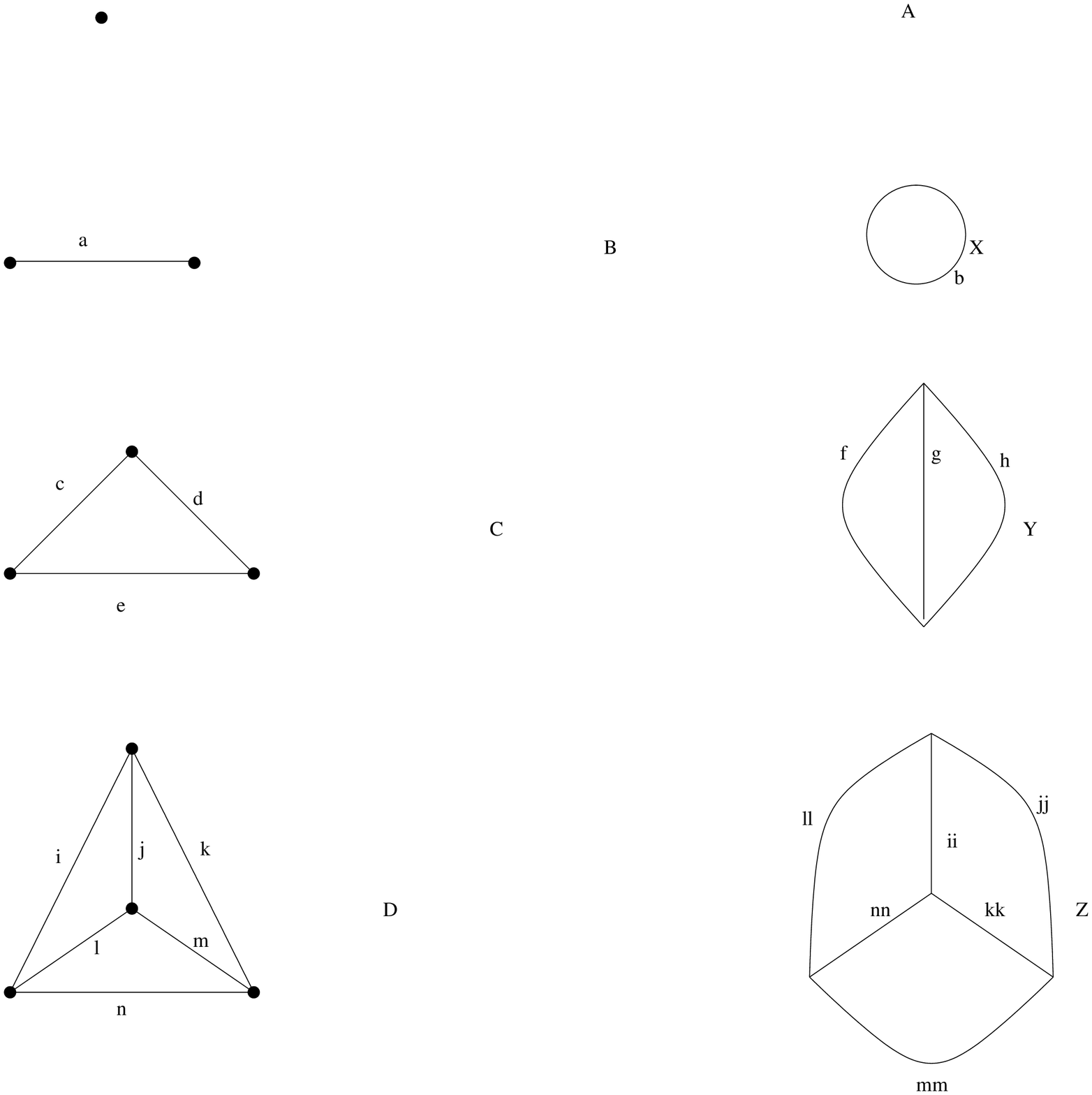}
\relabel{A}{${\n}^{-1}$}
\relabel{a}{$\s{a}$}
\relabel{B}{${\dim_q(a)=}\big  \langle $}
\relabel{b}{$\s{a}$}
\relabel{c}{$\s{c}$}
\relabel{d}{$\s{d}$}
\relabel{e}{$\s{e}$}
\relabel{f}{$\s{c}$}
\relabel{g}{$\s{d}$}
\relabel{h}{$\s{e}$}
\relabel{i}{$\s{i}$}
\relabel{j}{$\s{j}$}
\relabel{k}{$\s{k}$}
\relabel{l}{$\s{l}$}
\relabel{m}{$\s{m}$}
\relabel{n}{$\s{n}$}
\relabel{C}{${\theta(c,d,e)^{-1}=}\Big \langle$}
\relabel{ii}{$\s{i}$}
\relabel{jj}{$\s{j}$}
\relabel{kk}{$\s{k}$}
\relabel{ll}{$\s{l}$}
\relabel{mm}{$\s{m}$}
\relabel{nn}{$\s{n}$}
\relabel{D}{$\tau(i,j,k,l,m,n)= \Big \langle$}
\relabel{X}{$  \quad   \big \rangle$}
\relabel{Y}{$ \quad  \Big \rangle^{-1}$}
\relabel{Z}{$ \quad  \Big \rangle$}
\endrelabelbox }
\caption{\label{Weight} Weights associated with coloured simplices. All spin  networks are given the blackboard framing. }
\end{figure}

\begin{figure}
\centerline{\relabelbox 
\epsfysize 4cm
\epsfbox{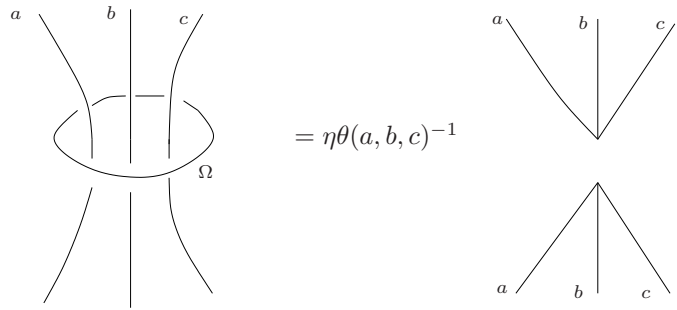}
\relabel{W}{$\s{\W}$}
\relabel{a}{$\s{a}$}
\relabel{b}{$\s{b}$}
\relabel{c}{$\s{c}$}
\relabel{aa}{$\s{a}$}
\relabel{bb}{$\s{b}$}
\relabel{cc}{$\s{c}$}
\relabel{aaa}{$\s{a}$}
\relabel{bbb}{$\s{b}$}
\relabel{ccc}{$\s{c}$}
\relabel{A}{$={\n}\theta(a,b,c)^{-1}$}
\endrelabelbox }\caption{\label{lic} Lickorish Encircling Lemma for the case of three strands. All networks are given the blackboard framing. }
\end{figure}
Using the identity shown in {Figure} \ref{lic} together with  {Figure} \ref{Tetrahedron}, it follows that:
\begin{align*}
Z_{\CH}(M)&=\sum_{\textrm{colourings of } M} \quad \prod_{\textrm{simplices } s \textrm{ of } M} \quad W(s)\\ &=Z_{\TV}(M).
\end{align*}
Note that we apply Lickorish Encircling Lemma to the 0-framed unknot defined from each face of the triangulation of $M$; see {Figure} \ref{arg}.  
The last expression for $Z_{\CH}$ is the usual definition of the Turaev-Viro Invariant.  For a complete proof of the fact  that $Z_{\TV}=Z_\CH$,  see \cite{R1}.
\begin{figure}
\centerline{\relabelbox 
\epsfysize 4cm
\epsfbox{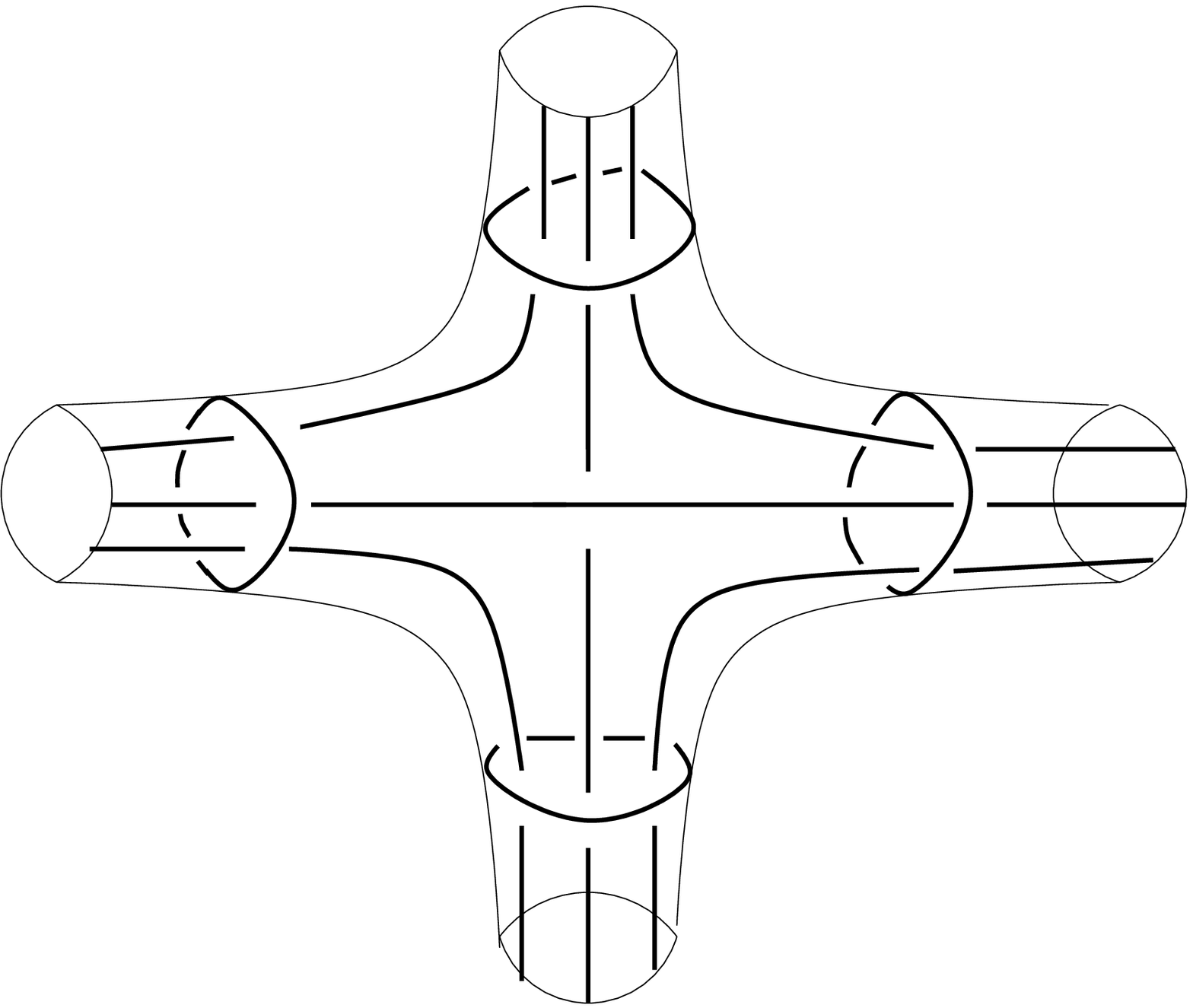}
\endrelabelbox }\caption{\label{Tetrahedron} Local configuration of the Chain-Mail Link at the vicinity of a tetrahedron. }
\end{figure}

\begin{figure}
\centerline{\relabelbox 
\epsfysize 2cm
\epsfbox{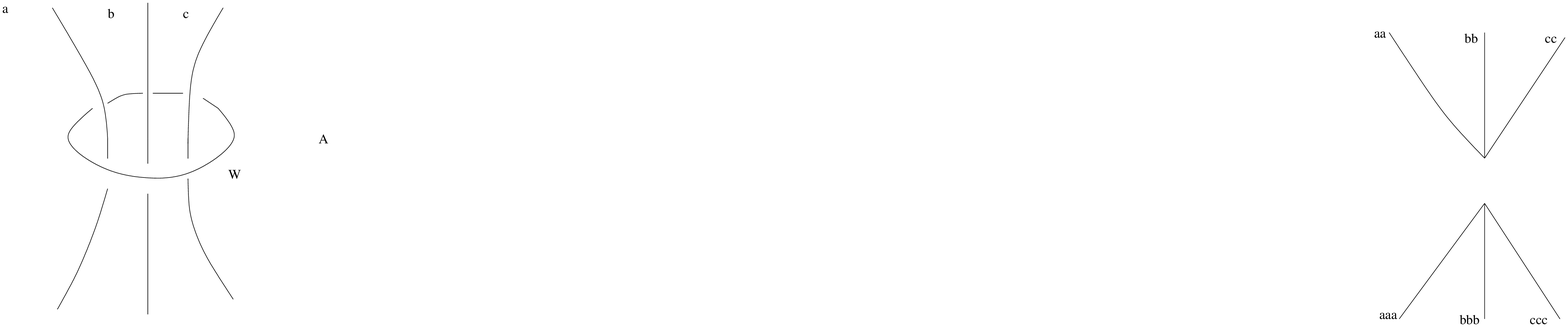}
\relabel{W}{$\s{\W}$}
\relabel{a}{$\s{\W}$}
\relabel{b}{$\s{\W}$}
\relabel{c}{$\s{\W}$}
\relabel{aa}{$\s{a}$}
\relabel{bb}{$\s{b}$}
\relabel{cc}{$\s{c}$}
\relabel{aaa}{$\s{a}$}
\relabel{bbb}{$\s{b}$}
\relabel{ccc}{$\s{c}$}
\relabel{A}{$=\displaystyle{{\n}\sum_{a,b,c} \dim_q(a) \dim_q(b) \dim_q(c)\theta(a,b,c)^{-1}}$}
\endrelabelbox }\caption{\label{arg} Applying Lickorish Encircling Lemma to the 0-framed link determined by a face of the triangulation of $M$. }
\end{figure}

\subsection{The Witten-Reshetikhin-Turaev Invariant}
The main references now are \cite{L}, \cite{RT} and \cite{R1}.
Let $M$ be an oriented {connected} closed $3$-manifold. Then $M$ can be
presented by surgery on some framed link $L \subset S^3$, up to orientation
preserving diffeomorphism. Any    {framed} graph $\G$ in $M$ can be pushed away from the areas where the surgery is performed, and therefore any pair $(M,\G)$, where $\G$ is a trivalent framed graph in the oriented closed  3-manifold $M$, can be presented as a pair $(L,{\G})$, where ${\G}$ is a framed trivalent graph in $S^3$, not intersecting $L$.    

The Witten-Reshetikhin-Turaev Invariant of a pair $(M,\G)$, where the framed graph $\G$  is coloured with the spins $j_1,...,j_n$,  is defined as:
$$Z_\WRT(M, \G;j_1,...,j_n)={\n}^{-\frac{\# L+1}{2}}\kappa ^{-\sigma(L)}\left <L \cup {\G};\W,j_1,...,j_n \right>.$$
Here $\sigma(L)$ is the signature of the linking matrix of the framed link $L$, and $\# L$ is the number of components of $L$.
This is an invariant of the pair $(M,\Gamma)$, up to orientation preserving diffeomorphism. In contrast with the Turaev-Viro Invariant, the  Witten-Reshetikhin-Turaev Invariant  is sensitive to the orientation of $M$. {If $M$ is an oriented 3-manifold, we represent the manifold with the reverse orientation by $\overline M$. }

With the normalisations that we are using, the Turaev-Walker theorem reads:
$$Z_\TV(M)=|Z_\WRT(M)|^2,$$
for any closed $3$-manifold $M$;
see \cite{R1,T3}.

Some other well known  properties of the  Witten-Reshetikhin-Turaev Invariant
are the following:
\begin{Theorem}\label{WRTp}
We have:
$$ Z_{\WRT}(S^3)={\n}^{-1/2}, \qquad Z_{\WRT}(S^2\times S^1)=1,$$
$$ Z_{\WRT}(\overline M)=\overline{Z_{\WRT}(M)},$$
$$    {Z_{\WRT}\big((P,\Gamma)\# (Q,\Gamma')\big )= Z_{\WRT}(P,\Gamma)Z_{\WRT}(Q,\Gamma'){\n}^{\frac{1}{2}}.}$$
\end{Theorem}
   {Here $M$, $P$ and $Q$ are oriented closed    {connected} 3-manifolds. In addition,   $\G$
  and $\G'$ are coloured graphs embedded in $P$ and $Q$.  It is understood
  that the connected sum $P \# Q$ is performed away from $\G$ and $\G'$.

Given oriented closed    {connected} 3-manifolds    {$P$ and $Q$}, we define    {$P\#_n Q$} in the following way; see \cite{BGM}. Remove $n$ 3-balls from    {$P$ and $Q$}, and glue the resulting manifolds    {$P'$ and $Q'$} along their boundary in the obvious way,    {so that the final result is an oriented manifold.}    {We denote it by:} $$   {P\#_n Q= P' \bigcup_{\d P' = \d Q'} Q'.}$$ 
 {It is immediate that}    {$P \#_1 Q=P \# Q$,} and {that}    {$P \#_n Q$} is diffeomorphic to    {$(P \# Q)\# (S^1 \times S^2)^{\# (n-1)}$, if $n>1$.} By using Theorem \ref{WRTp} it follows that: 
\begin{multline}\label{refer} 
Z_\WRT(P \#_n Q, \Gamma \cup \Gamma';j_1,...,j_p,i_1,...,i_m)\\=   
Z_\WRT(P, \Gamma;j_1,...,j_p)Z_\WRT( Q,  \Gamma';i_1,...,i_m)\n^{{\frac{n}{2}}}.
\end{multline}
 Here $P$ and $Q$ are closed oriented 3-manifolds. In addition, $\G$ and $\G'$
 are graphs in $P$ and $Q$, coloured with the spins $j_1,...,j_p$ and
 $i_1,...,i_m$, respectively. As before, it is implicit that the multiple
 connected sum $P \#_n Q$ is performed away from $\G$ and $\G'$. 

\section{Perturbative expansion}

{In this section we are going to define the $Z_n$'s considered in the introduction.}

\subsection{Graspings}\label{grasping}
Let ${\rm K}$ be a simplicial complex whose geometric realisation $|{\rm K}|$ is a piecewise-linear closed $p$-dimensional manifold. Recall that we can define the dual cell decomposition of $|{\rm K}|$, where each $k$-simplex of ${\rm K}$ generates a dual $(p-k)$-cell of the dual cell decomposition of $|{\rm K}|$, see \cite[3.3]{TV}, for example. This is very easy to visualise in three dimensions.

\begin{Definition}[$n$-grasping]
Let ${\T}\subset \R^3$ be the standard tetrahedron.  For a positive integer $n \in \N$, an $n$-grasping $\g$ is a sequence $(v_1v_2\ldots v_n)$, where $v_k$ is a vertex of $\T$ for each $k =1,2,\ldots,n$.
\end{Definition}

Any $1$-grasping $\g=(v)$, where $v$ is a vertex of $\T$, naturally  defines a   trivalent graph $G_\g$ on the boundary $\d \T$ of $\T$ (usually called a grasping itself), by doing the transition shown in {Figure} \ref{grasp}. The graph $G_\g$ is therefore the union $G_\T \cup \i_v$, where $G_\T$ is the dual graph to the 1-skeleton of the obvious triangulation of $\d \T$, and $\i_v$ is homeomorphic to the graph ${\rm Y}$ made from a trivalent vertex and three open-ends.

\begin{figure}
\centerline{\relabelbox 
\epsfysize 4cm
\epsfbox{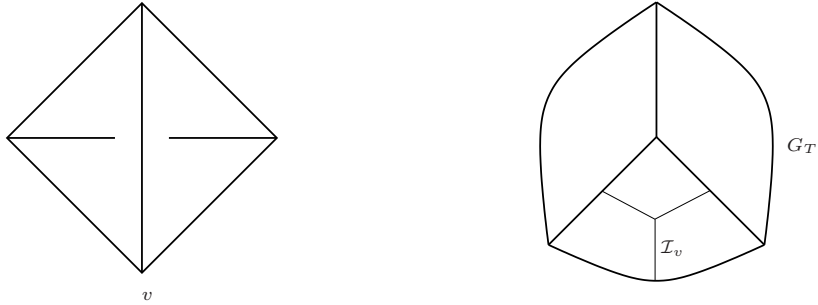}
\relabel{v}{$\s{v}$}
\relabel{i}{$\s{\i_v}$}
\relabel{G}{$\s{G_T}$}
\endrelabelbox } 
\caption{\label{grasp} The form of the graph $G_\g=G_\T \cup \i_v\subset \d \T$ defined from a grasping $\g=(v)$, where $v$ is a vertex of $\T$. Note that $\i_v$ is placed inside the dual face to $v$, and thus it intersects $G_\T$ in the dual edges of the edges incident to $v$.}
\end{figure}

We want to define, in an analogous fashion, a trivalent graph $G_\g\subset \d \T$ from an $n$-grasping  $\g=(v_1v_2 \ldots v_n)$. This is not possible unless further information  is given. We  want $G_\g$ to be the union of $G_\T$ and a disjoint union $\sqcup_{i=1}^n \Y_i$, where each graph $\Y_i$ is homeomorphic to the graph $\Y$. To describe $G_\g$ we need to specify where the ends of each $\Y_i$ intersect $G_\T$, as well as the crossing information. To this end we give the following definition:

\begin{Definition}[Space ordering of an $n$-grasping]
Let again $\T\subset \R^3$ be the standard tetrahedron. Let $G_\T$ be the dual graph to the obvious triangulation of the boundary $\d \T$ of $\T$. Any edge $e$ of $\T$ therefore defines a dual edge $e^*$ of the graph $G_\T\subset \d \T$. Let $\g=(v_1\ldots v_n)$ be an $n$-grasping.  A space pre-ordering of $\g$ is given by an assignment of a subset {$O^i=\{x^i_1,x^i_2,x^i_3\} \subset G_\T$} to each $i =1,2,\ldots, n$ such that:
\begin{enumerate}
 \item For each $i$, $x^i_1,x^i_2$ and $x^i_3$ belong to different  edges of $G_\T$, and each of these points belongs to the dual edge of an edge incident to $v_i$.
\item $O^i \cap O^j = \emptyset$ if $i \neq j$.
\end{enumerate}
A space ordering $\O_\g$ of $\g$ is given by a space pre-ordering of $\g$ considered up to {ambient} isotopy of $\cup_{i=1}^n O^i$ inside $G_\T$.
\end{Definition}
There exists therefore a unique space ordering of a 1-grasping $\g=(v)$.

Let now $\g=(v_1v_2\ldots v_n)$ be an $n$-grasping with a certain space ordering $\O_\g=\left(\{x^i_1,x^i_2,x^i_3\}\right)_{i=1}^n$. We define an associated graph $G(\g,\O)$ in $\d \T$, with crossing information, as being:
$$G(\g,\O)=G_\T \cup \bigcup_{i=1}^n \i^i_{v^i} ,$$ 
where:
\begin{enumerate}
\item  $\i^i_{v^i}$ is homeomorphic to $\i_{v^i}$ (see above) for each $i=1,\ldots n$.
 \item $\i_{v_i}^i \cap G_\T=\{x^i_1,x^i_2,x^i_3\}$ for each $i=1,2,\ldots,n$.
\item If $i <j$ then $\i_{v_i}^i$ is placed above $\i_{v_j}^i$, with respect to the boundary of $\T$.
\end{enumerate}
See {Figure} \ref{grasp2} for the description of the graph $G(\g,\O_\g)$ for two different space orderings of $\g=(vvw)$, where $v \neq w$. 

\begin{figure}
\centerline{\relabelbox 
\epsfysize 3.4cm
\epsfbox{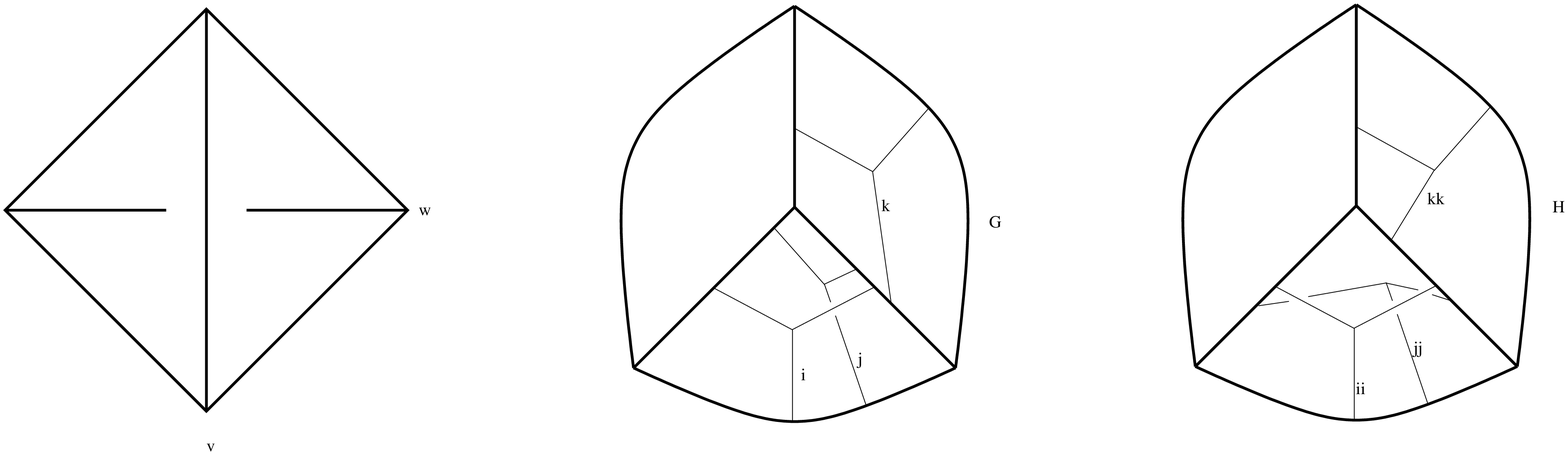}
\relabel{v}{$\s{v}$}
\relabel{w}{$\s{w}$}
\relabel{i}{$\s{\i_v^1}$}
\relabel{j}{$\s{\i_v^2}$}
\relabel{k}{$\s{\i_w^3}$}
\relabel{ii}{$\s{\i_v^1}$}
\relabel{jj}{$\s{\i_v^2}$}
\relabel{kk}{$\s{\i_w^3}$}
\relabel{G}{$\s{G_T}$}
\relabel{H}{$\s{G_T}$}
\endrelabelbox } 
\caption{\label{grasp2} Defining the graph $G(\g,\O_\g)$ defined from a grasping $\g=(vvw)$ for two different space orderings of $\g$.}
\end{figure}

{Given a $1$-grasping $\g=(v)$ living in the {tetrahedron $\T$,} define $\sigma(v)$ as being given by the set made from the three edges of $\T$ incident {to $v$.}
Given an $n$-grasping $\g=(v_1\ldots v_n)$ living in $\T$, define {$\sigma(\g)=\cup_{i=1}^n \sigma(v_i)$.}}

\subsection{The {first-order} volume expectation value}\label{vev}
Let $M$ be a piecewise linear oriented closed 3-manifold (from now on called simply a $3$-manifold). Consider a triangulation {$\D$} of $M$. Let $M_0,M_1,M_2$ and $M_3$ be the set of vertices, edges, triangles and tetrahedra of $M$.  

A colouring of $M$ is an assignment of a spin    {$j \in \{0,1/2,...,(r-2)/2\}$} to each edge of $M$. Each colouring of a simplex $s$ of $M$ gives rise to a weight $W(s)$, in the way shown in {Figure} \ref{Weight}, exactly the same fashion as in the definition of the Turaev-Viro Invariant $Z_\TV$.

 Consider a tetrahedron $T$ of $M$ (whose edges are coloured), with some  $n$-grasping $\g=(v_1\ldots v_n)$, provided with  a space ordering $\O$. Choose an orientation preserving embedding of $T$ into $S^3$, which is defined up to isotopy. Then the weight $W(T,\g,\O)$ is defined as being {the factor ${A}(T,\g)$ (see below)} times the evaluation of the spin network $G(\g,\O)=G_T\cup \bigcup_{i=1}^n \i^i_{v^i} \subset \d T \subset S^3$, where $G_T$ has the colouring given by the colouring of $M$ (recall that each edge of $G_T$ is dual to a unique edge of $T$), and all edges of each $\i^i_{v^i}$ are assigned the spin $1$. Note that the graph  $G(\g,\O)\subset \d\T$ has a natural framing parallel to the boundary of $T$. {If $\g=(v_1,\ldots,v_n)$ is an $n$-grasping, the factor ${A}(T,\g)$ is, by definition:}
$${\prod_{i=1}^n {A}_{l_1^i}{A}_{l_2^i}{A}_{l_3^i},} $$
{where for each $i$ we let $l_1^i$,  $l_2^i$ and $l_3^i$ denote the colourings of the three edges of $T$ incident to $v_i$ and} {the $A_l$'s are given by (\ref{aj}).}
 
The {first-order} volume expectation value {can be represented as
$$ \langle V(M) \rangle = \int {\cal D} A \,{\cal D} B\, V(M)\, e^{i\int_M Tr(B\wedge F)}\,,$$
where $V(M) =\int_M \epsilon_{abc}B^a \wedge B^b \wedge B^c $ is the volume of $M$.}
It corresponds to $-iZ_1 = i\sum_{k=1}^N \partial_J^3(\tau_k) Z_J {\Big|}_0$, {which} can be defined as
\begin{equation}{i\V(M,\D)}= {\frac{i}{4}}
 \sum_{T \in M_3}\quad  \sum_{\textrm{1-graspings } \g \textrm{ of } T} \v(M,T,\g),
\end{equation}
where, by definition:
\begin{multline}
\v(M,T,\g)= \sum_{\textrm{colourings of } M } \quad W\big (T,\g,\O)\\ \prod_{T' \in M_3 \setminus \{T\}} W(T') \prod_{s \in M_0 \cup M_1 \cup M_2} W(s).
\end{multline}
Recall that any $1$-grasping $\g$ has a unique space ordering $\O$. 

Let $v(M)={\V(M,\D)/N}$, where $N$ is the number of tetrahedra of $M$.
\begin{Theorem}\label{volc} {For any triangulation of $M$ we have 
$v(M) =0$.} 
\end{Theorem}

\begin{Proof}
Each term $\v(M,T,\g)$ can be presented in a Chain-Mail way. As in
\cite{R1}, consider the natural  handle decomposition of $M$ for which each
$i$-simplex of $M$ generates a $(3-i)$-handle of $M$. This handle
decomposition is dual to the one where each $i$-simplex of $M$ is thickened to
an $i$-handle of $M$.

Let us consider the Chain-Mail formula $$Z_\CH(M)={\n}^{-n_0-n_2}\left <
  \CH(H_-,m,\e,\Phi);\W \right>,$$ for $Z_\TV(M)$, obtained from this  handle
decomposition of $M$; see \ref{TViro} and \cite{R1}.  Here $n_i$ is the number of $i$-handles of $M$, and therefore it equals the number of $(3-i)$-simplices of $M${. From} the same argument that shows that $Z_\TV(M)=Z_\CH(M)$, follows that:
\begin{multline*}
\v(M,T,\g)=\sum_{a,b,c }{{A}_a {A}_b {A}_c} \dim_q(a) \dim_q(b) \dim_q(c)  {\n}^{-n_0-n_2}\\\left <\CH(H_-,m ,\e_{T,\g}, (l_1 \sqcup l_2\sqcup l_3)\cup \Y_{v},\Phi);\W,\W,a,b,c,1\right>,
\end{multline*}
where:
\begin{enumerate}
 \item All components of $m$ are coloured with $\W$.
\item Recall that each circle of the link $\e$ corresponds to a certain edge
  of $M$. The components $l_1,l_2$ and $l_3$  of $\e$ which correspond to the
  edges $e_1$, $e_2$ and $e_3$ incident to $v$, where $\g=(v)$, should be
  coloured with the spins $a,b,c \in  \{0,1/2,...,(r-2)/2\}$, whereas the remaining components (which form the link $\e_{T,\g}$) should be coloured with $\W$.
\item The component $\Y_v$, where $\g=(v)$, is a trivalent vertex with three open ends, each of
  which is incident to either $l_1$, $l_2$ or $l_3$, with no repetitions, with framing parallel to the surface of $H_-$; see {Figure} \ref{chainins}. The three edges of $\Y_v$ are to be assigned the spin $1$.
\item Finally, $\Phi$ is an orientation preserving embedding $H_-\to S^3$. As in the case where no graspings are present, the final result is independent of this choice; see \cite[Proof of Proposition 3.3]{R1}.
\end{enumerate}
\begin{figure}
\centerline{\relabelbox 
\epsfysize 6cm
\epsfbox{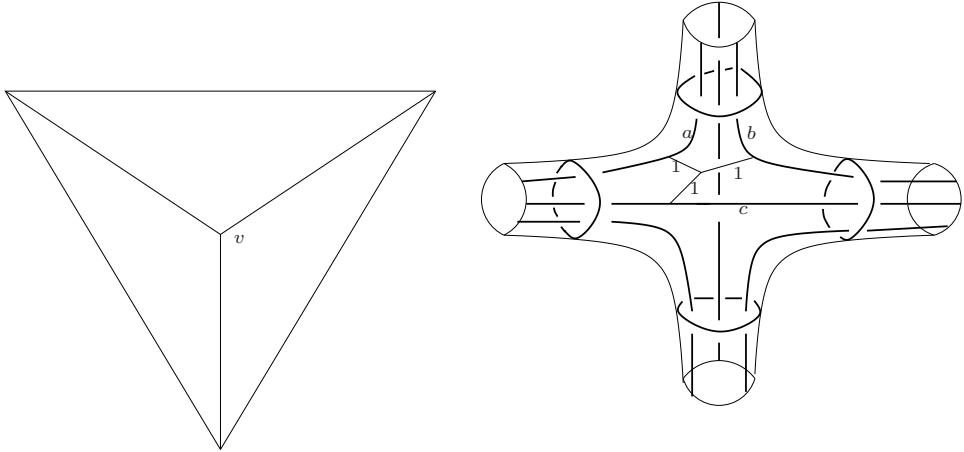}
\relabel{v}{$\s{v}$}
\relabel{1}{$\s{1}$}
\relabel{2}{$\s{1}$}
\relabel{3}{$\s{1}$}
\relabel{a}{$\s{a}$}
\relabel{b}{$\s{b}$}
\relabel{c}{$\s{c}$}
\endrelabelbox }\caption{\label{chainins} Local configuration of the Chain-Mail Link at the vicinity of a tetrahedron $T$, for the case when  $T$ has a grasping $\g=(v)$. All strands are coloured with $\W$, unless indicated.}
\end{figure}

By cancelling some pairs of $0$- and $1$-handles, we can reduce the handle
decomposition of $M$ to one with a single $0$-handle. Similarly, by
eliminating pairs of $2$- and $3$-handles, we can reduce the handle
decomposition of $M$  to one having four $3$-handles, each of which corresponds to one of the vertices of the triangulation of $M$ which are endpoints of the edges of $T$ incident to $v$, where $\g=(v)$, and  so that the   2-handles
corresponding to the three edges of $T$ incident to $v$ are still in the handle
decomposition.  The  chain-mail link of the new handle decomposition of $M$ will
then be  $\CH(H_-,m',\e',\Phi)$, where $m'$ is obtained from $m$ by removing
some circles, and the same for $\e'$. Let $n_i'$ be the number of $i$-handles
of the new handle decomposition of $M$.

Let $\e_{T,\g}'=\e'\setminus \{l_1 \sqcup l_2\sqcup l_3\}$.
 By the same argument as in  \cite[Proof of Theorem 3.4]{R1}  follows:
\begin{multline*}
{v_\TV(M,T,\g)}=\sum_{a,b,c }{{A}_a {A}_b {A}_c} \dim_q(a) \dim_q(b) \dim_q(c)  {\n}^{-n_0'-n_2'}\\ \left <\CH(H_-,m' ,\e_{T,\g}', (l_1 \sqcup l_2\sqcup l_3)\cup \Y_{v},\Phi);\W,\W,a,b,c,1 \right>.
\end{multline*}

Given a compact $3$-manifold with border  $Q$ embedded in the oriented 3-manifold  $M$,  define $M\#_Q {\overline{M}}$
as being the manifold obtained from $M$ and ${\overline{M}}$ by removing the interior of $Q$ from each of
them  and gluing the resulting manifolds along the identity map $\d (M \setminus Q )\to \d({\overline{M \setminus Q}})$.

Let $S_\g$ be the graph in $M$ made from  the {edges of $T$ incident to $\g=(v)$,} together with their endpoints. Each edge of $S_\g$ will induce a $2$-handle of $M$
and its four vertices  will induce  $3$-handles of $M$. The union of these
handles will be a regular neighbourhood ${\rho(}S_\g)$ of $S_\g$. Consider the graph $R_\g$ in $\d {\rho(}S_\g)$ made from the attaching spheres of these $2$-handles, with a $\Y$-graph  inserted, as in {Figure} \ref{graph}.
\begin{figure}
\centerline{\relabelbox 
\epsfysize 5cm
\epsfbox{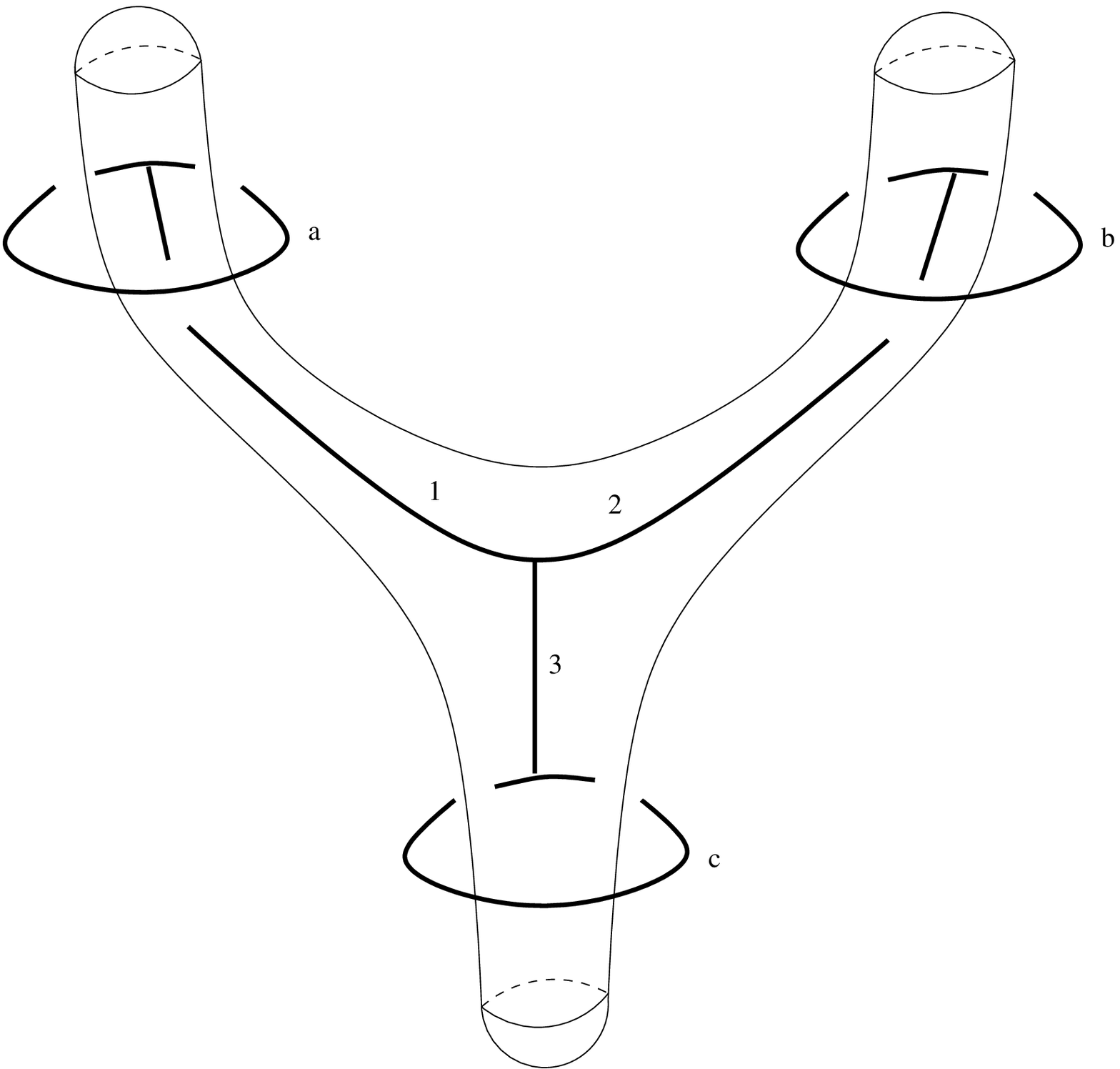}
\relabel{1}{$\s{1}$}
\relabel{2}{$\s{1}$}
\relabel{3}{$\s{1}$}
\relabel{a}{$\s{a}$}
\relabel{b}{$\s{b}$}
\relabel{c}{$\s{c}$}
\endrelabelbox }\caption{\label{graph} The graph $R_\g$ inside $\d {\rho(}S_\g)$.}
\end{figure}
\begin{figure}
\centerline{\relabelbox 
\epsfysize 4cm
\epsfbox{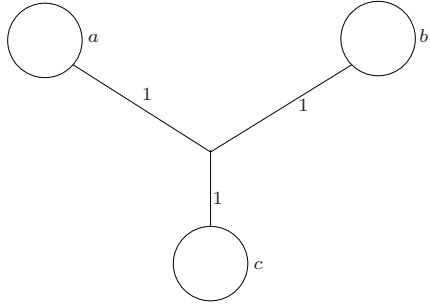}
\relabel{1}{$\s{1}$}
\relabel{2}{$\s{1}$}
\relabel{3}{$\s{1}$}
\relabel{a}{$\s{a}$}
\relabel{b}{$\s{b}$}
\relabel{c}{$\s{c}$}
\endrelabelbox }\caption{\label{haty} The coloured graph ${\hat{\Y}}$.}
\end{figure}

By the using either the  argument in \cite[Proof of Theorem 1]{BGM}) or in  \cite[Proof of Lemma 3.3]{FMM},
the pair $(\Phi(m' \cup \e_{T,\g}'), \Phi((l_1 \sqcup l_2\sqcup l_3)\cup
\Y_{v}))$ is a surgery presentation of the manifold $M\#_{{\rho(}S_\g)}{\overline{M}}$, with
the graph $R_\g \subset \d {\rho(}S_\g) $ inserted in $\d (M \setminus {\rho(}S_\g))$. 

The signature of the link $\Phi(m' \cup \e_{T,\g}')$ is zero, given that it is a Kirby diagram for the manifold $(M\setminus {\rho(}S_\g)) \times I$; see \cite[Proof of Theorem 3.7]{R1} or \cite[Proof of Theorem 1]{BGM}. Therefore it follows that:
\begin{align*}
\v(M,T,\g)&=\sum_{a,b,c }{{A}_a {A}_b {A}_c}\dim_q(a)\dim_q(b)\dim_q(c) \\ & \quad \quad \quad {\n}^{-\frac{7}{2}}  Z_\WRT(M\#_{{\rho(}S_\g)}{\overline{M}},R_\g;a,b,c,1).
\end{align*}
We have used the calculation $\n^{-n_0'-n_2'+\frac{1+n_1'+n_2'-3}{2}}=\n^{-\frac{7}{2}}$, which follows from the fact that the Euler characteristic of a  closed orientable $3$-manifold is zero. Note that $n_0'=1$ and $n_3'=4$, by construction.

Since ${\rho(}S_\g)$ is a closed $3$-ball embedded in $M$ {it} follows that
$M\#_{{\rho(}S_\g)}{\overline{M}} \cong M\#{\overline{M}}$. On the other hand, the graph $R_\g$ is
obviously trivially embedded in $M \#{\overline{M}}$, in the sense that there exists an
embedding  $B^3 \to M$  sending the graph ${{\hat{\Y}}}$ {of Figure \ref{haty}} to $R_\g$. This leads to:
\begin{align*}
&\v(M,T,\g)\\&=\sum_{a,b,c }{{A}_a {A}_b {A}_c}\dim_q(a)\dim_q(b)\dim_q(c) {\n}^{-\frac{7}{2}} Z_\WRT(M\#_{{\rho(}S_\g)}{\overline{M}},R_\g;a,b,c,1)\\
&=\sum_{a,b,c }{{A}_a {A}_b {A}_c}\dim_q(a)\dim_q(b)\dim_q(c) {\n}^{-\frac{7}{2}} Z_\WRT(M\#{\overline{M}}\#S^3,{{\hat{\Y}}};a,b,c,1)\\
&=\sum_{a,b,c}{{A}_a {A}_b {A}_c}\dim_q(a)\dim_q(b)\dim_q(c) {\n}^{-\frac{7}{2}+1} Z_\TV(M)Z_\WRT(S^3,{{\hat{\Y}}};a,b,c,1)\\
&=\sum_{a,b,c}{{A}_a {A}_b {A}_c} \dim_q(a)\dim_q(b)\dim_q(c){\n}^{-3}  Z_\TV(M)\left <{{\hat{\Y}}};a,b,c,1\right>.
\end{align*}
Since {the graph $ {{\hat{\Y}}}$ consists of three tadpole graphs, and the evaluation of a tadpole quantum spin network is zero, see {\cite[Theorem 3.7.1]{CFS} or \cite[Chapter 9]{KL},} then $<{{\hat{\Y}}};a,b,c,1> = 0$, which implies $v(M)=0$.}
\end{Proof}

{Given that   $Z_1  = - \V$, we have}
$$ Z_1(M,\Delta) = N z_1 Z_\TV(M) = 0\,,$$
{where $$ z_1 =-{\n}^{-3}\sum_{a,b,c} {A}_a {A}_b {A}_c \dim_q(a)\dim_q(b)\dim_q(c) \left <{{{\hat{\Y}}}};a,b,c,1\right> =0 \,,$$
{and $N$ is the number of tetrahedra of $M$.}
\subsection{Higher-order corrections}\label{Hoc}

Let $M$ be a 3-manifold with a triangulation $\D$. Let $n$ be a positive integer. An $n$-grasping $\Gc$ of $M$ is a set $\Tc_\Gc=\{T_1^\Gc,\ldots, T_{m_\Gc}^\Gc\}$ of tetrahedra of $M$ (where $T_i^\Gc\neq T_j^\Gc$ if $i\neq j$), each of which is provided with a space ordered $n_i^\Gc$-grasping $(\g_i^\Gc,\O_i^\Gc)$, where $n_i^\Gc>0$, such that $n_1^\Gc+n_2^\Gc+\ldots+n_{m_\Gc}^\Gc=n$. The set $\Tc_\Gc$ is said to be an $n$-grasping support and the $n$-grasping $\Gc$ of $M$ is said to be supported in $\Tc_\Gc$.

Recall the definition of the weights $W(T,\g,\O)\in \C$, where $T$ is a coloured tetrahedra, with a space ordered  grasping $(\g,\O)$ living in $T$. This appears in the beginning of Subsection \ref{vev}, to which we refer for the notation below. 

Define:
\begin{multline}\label{ex1}
{\V^{(n)}(}M,\D)= {\frac{1}{4^n}}\sum_{n\textrm{-graspings } \Gc \textrm{ of } M}  \quad \sum_{\textrm{colourings of } M } \quad \\ \prod_{i=1}^{m_\Gc}\quad W(T^\Gc_i,\g^\Gc_i,\O^\Gc_i) \prod_{T' \in M_3 \setminus \Tc_\Gc} W(T') \prod_{s \in M_0 \cup M_1 \cup M_2} W(s),
\end{multline}
which can also be written as
\begin{multline}\label{ex2}
{\V^{(n)}(}M,\D)= {\frac{1}{4^n}}\sum_{K=1}^n \quad \sum_{\substack{ n-\textrm{graspings supports }  \Tc\\ \textrm{with } K \textrm{ tetrahedra}}}\quad \sum_{\substack{n-\textrm{graspings } \Gc \textrm{ of } M \\ \textrm{supported in } \Tc}} \\ \sum_{\textrm{colourings of } M } \quad \prod_{i=1}^{K}\quad W(T^\Gc_i,\g^\Gc_i,\O^\Gc_i) \prod_{T' \in M_3 \setminus \Tc} W(T') \prod_{s \in M_0 \cup M_1 \cup M_2} W(s).
\end{multline}

Observe that $\V^{(n)}$ is related to the expectation value of the $n$-th power of the volume $V=\int_M \epsilon_{abc} B^a \wedge B^b \wedge B^c $:
$$\langle V^n \rangle \equiv \hat{V}^n Z(J){\Big |}_0 = i^n n! \V^{(n)}\,,$$
where $\hat V = i \sum_{k=1}^N \d_J^3 (\tau_k)$ is the volume operator. Furthermore
\begin{equation} 
Z=\sum_{n=0}^\infty \frac{i^n\lambda^n}{n!}\langle V^n \rangle = \sum_{n=0}^\infty (-1)^n \lambda^n\V^{(n)} \,.
\label{pev}\end{equation}

{Let us analyse whether expressions (\ref{ex1}) and (\ref{ex2}) define a topological invariant. Consider the bottom term of (\ref{ex2}):}
\begin{multline*}
X(M,\D,K,\Gc)\\= \sum_{\textrm{colourings of } M } \quad \prod_{i=1}^{K}\quad W(T^\Gc_i,\g^\Gc_i,\O^\Gc_i) \prod_{T' \in M_3 \setminus \Tc_\Gc} W(T') \prod_{s \in M_0 \cup M_1 \cup M_2} W(s).
\end{multline*}
It depends on an $n$-grasping $\Gc=\{T_i^\Gc,\g_i^\Gc,\O_i^\Gc\}_{i=1}^K$ of $M$ supported in the set with $K$ tetrahedra $\Tc_\Gc=\{T_i^\Gc\}_{i=1}^K$. 

As in the proof of Theorem \ref{volc},   the value of $X(M,\D,K,\Gc)$ can  be presented as the evaluation of a chain-mail link, with some additional 3-valent vertices inserted. 

{Let $\Tc^1_\Gc$ be given by}
\begin{equation}\label{F}
{\Tc^1_\Gc=\bigcup_{i=1}^K \sigma\left ( \g_i^\Gc \right),}
\end{equation}
{see the end of Subsection \ref{grasping} for this notation.  Let also }
\begin{equation}\label{GG}{{A}(\Gc,c)=\prod_{i=1}^K A\big(T_i^\Gc,\g^\Gc_i\big);}
\end{equation}
{depending on a colouring $c$ of $M$; see subsection \ref{vev}.}

We have
\begin{multline}
X(M,\D,K,\Gc)\\= \sum_{\textrm{colourings } c \textrm{ of } \Tc^1_\Gc }{A(\Gc,c)} \dim_q(c){\n}^{-n_0-n_2}\left <\CH(m,\e_{\Tc_\Gc},L_\Gc,\Phi);\W,c,1\right>,
\end{multline}
where:
\begin{enumerate}
 \item All components of {the link} $m$ are coloured with $\W$.
\item The graph $L_\Gc$ is made from the attaching regions of the 2-handles of $M$ which correspond to the edges of the triangulation of $M$ belonging to  $\Tc_\Gc^1$, with the obvious colouring,  with $n$  $\Y$-graphs inserted in the obvious way, and coloured by the spin-$1$ representation.
\item The link $\e_{\Tc_\Gc}$ is formed by the attaching regions of the 2-handles of $M$ corresponding to the remaining edges of $M$. These should be coloured with the  $\W$-element.
\item We have  put $$\dim_q(c)=\prod_{e \in \Tc^1_\Gc}\dim_q c(e).$$ 
\item As usual $\Phi$ is an embedding $H_- \to S^3$. The final result is independent of this {choice}.
\end{enumerate}

We can now reduce the handle decomposition of $M$ to one with a unique $0$-handle, and so that all $n_3'$  $3$-handles of it are dual to vertices  of $M$ occurring as endpoints of edges in  $\Tc_\Gc^1$.  {Moreover, we can suppose that the $2$-handles of $M$ which are dual to the edges of $\Tc^1_\Gc$ are still in the new handle decomposition.} Let ${\rho(}\Tc^1_\Gc)$ be a regular neighbourhood of $\Tc^1_\Gc$ in $M$.  Similarly to the $n=1$ case, the graph $L_\Gc$ {naturally projects} to a graph $R_\Gc$ in $\d({\rho(}\Tc^1_\Gc)),$ with crossing information; see Figure \ref{example} for an example. By the same argument as in the proof of Theorem \ref{volc} {it} follows  that:
\begin{multline}\label{xm}
X(M,\D,K,\Gc)=\\ 
\sum_{\textrm{colourings } c \textrm{ of } \Tc^1_\Gc }{A(\Gc,c)} \dim_q(c){\n}^{-\frac{n_3'}{2}-\frac{\# \Tc^1_\Gc}{2}} Z_\WRT\left(M \#_{{\rho(}\Tc^1_\Gc)} {\overline{M}},R_\Gc\right).
\end{multline}

\begin{figure}
\centerline{\relabelbox 
\epsfysize 6cm
\epsfbox{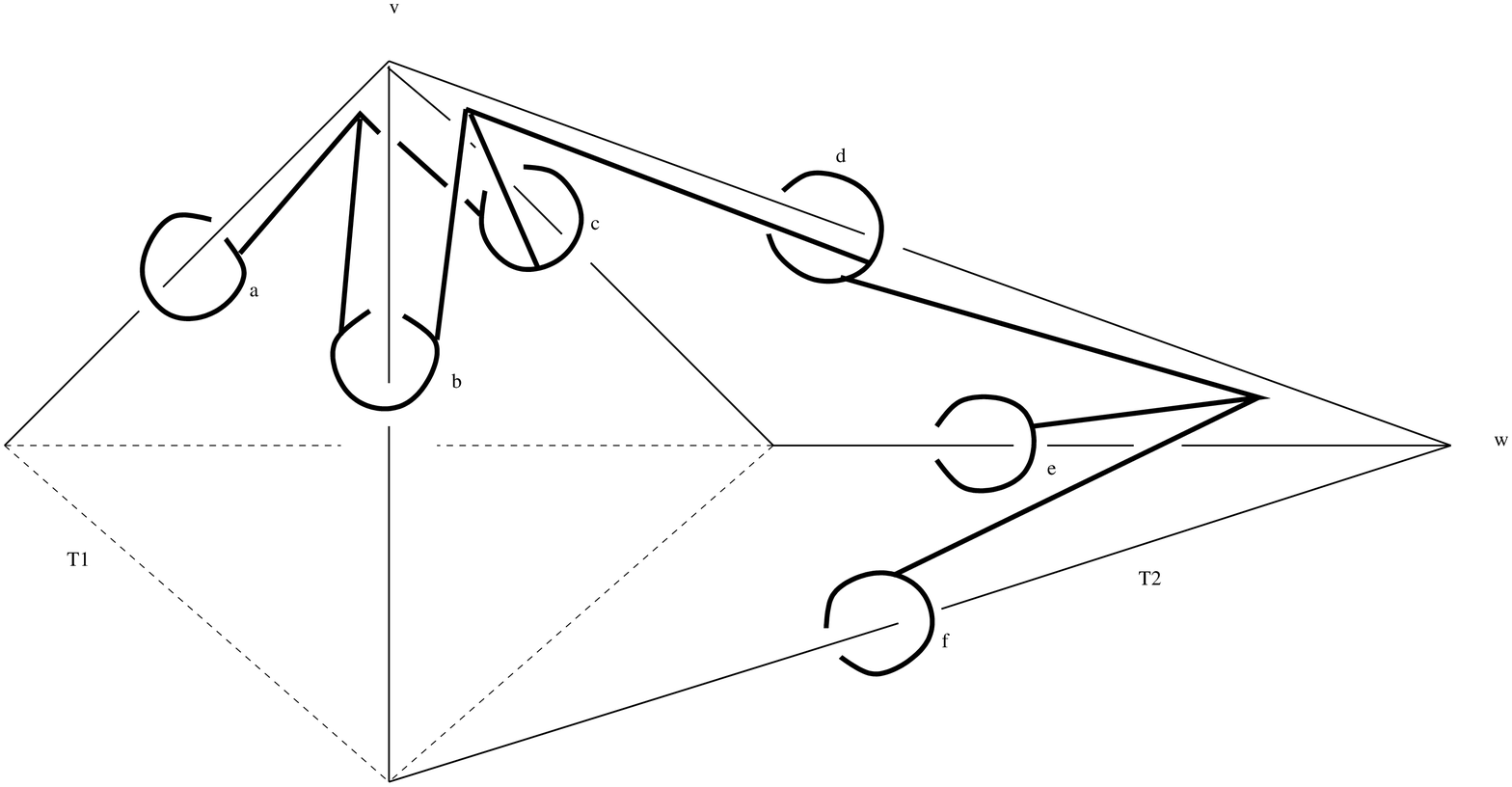}
\relabel{v}{$\s{v}$}
\relabel{w}{$\s{w}$}
\relabel{T1}{$\s{T_1}$}
\relabel{T2}{$\s{T_2}$}
\relabel{a}{$\s{a}$}
\relabel{b}{$\s{b}$}
\relabel{c}{$\s{c}$}
\relabel{d}{$\s{d}$}
\relabel{e}{$\s{e}$}
\relabel{f}{$\s{f}$}
\endrelabelbox }\caption{\label{example} The graph $R_\Gc$ in $\d({\rho(}\Tc^1_\Gc))$. Here $\Gc$ is a grasping of $M$ supported in  the set $\Tc^1_\Gc=\{T_1,T_2\}$,  with $\g^1_\Gc=(v)$ and $\g^2_\Gc=(vw)$, and the  space ordering shown. In this example, ${\rho(}\Tc^1_\Gc)$ is obtained by thickening the solid edges of $T_1$ and $T_2$. }
\end{figure}

In contrast to the case when $\Gc$ is a 1-grasping, the expression (\ref{xm}) is not apriori a topological invariant of $M$. This is because there can
{exist} several subsets of $M$ that are of the form ${\rho(}\Tc^1_\Gc)$, {for some $n$-graping   $\Gc$ of $M$ with $K$ tetrahedra, if we consider an arbitrary triangulation $\D$ of
$M$;} we will go back to this later. This is the reason why a similar result to Theorem \ref{volc} does not
immediately hold for ${{\V^{(n)}}(}M,\D)$ for $n>1$. 

Note that equation (\ref{xm}) simplifies to
\begin{align} 
&X(M,\D,K,\Gc) \nonumber\\
& =\sum_{\textrm{colourings } c \textrm{ of } \Tc^1_\Gc }
{A(\Gc,c)}\dim_q(c){\n}^{-\frac{n_3'}{2}-\frac{\# \Tc^1_\Gc}{2}  } Z_\WRT\left(M  \# (S^3\#_{{\rho(}\Tc^1_\Gc)} {\overline{S^3}}) \#
  {\overline{M}},R_\Gc\right)\\\label{x2}
&  =\sum_{\textrm{colourings } c \textrm{ of } \Tc^1_\Gc }{A(\Gc,c)}
\dim_q(c){\n}^{1-\frac{n_3'}{2}-\frac{\# \Tc^1_\Gc}{2} } Z_\TV(M)Z_\WRT\left(S^3\#_{{\rho(}\Tc^1_\Gc)} {\overline{S^3}},R_\Gc\right),
\end{align}
whenever $\Tc^1_\Gc$  is confined to a closed ball contained in $M$. For fixed $n$, this happens whenever the triangulation  $\D$ of $M$ is fine enough.

\section{Dilute Gas Limit}\label{Dilute}

Let $M$ be a  3-manifold. Similarly to \cite{Ba2},  to eliminate the triangulation dependence of ${\V^{(n)}(}M,\D)$, we want to consider  the limit 
\begin{equation}\label{Lim}
\lim_{|\D|\to 0}\frac{1}{N^n}{\V^{(n)}(}M,\D),
\end{equation} in a sense that still needs to be addressed.  Here $N=N_\D$ denotes
the number of tetrahedra of a triangulation $\D$  of $M$.
The case considered in \cite{Ba2} is the limit when the maximal diameter of
each tetrahedra of a triangulation $\D$ of $M$ tends to zero, called there the
``{\it Dilute Gas Limit.}''

\subsection{Preliminary approach}\label{pre}

\noindent{\bf Warning} {\it There exists a gap in the argument below; see Assumption 1. It corresponds to Conjecture $B$ in page 8 of \cite{Ba2}. In Subsection \ref{exact} we explain how we can go  around it by restricting the class of triangulations with which we work, so that all calculations are valid.}

\quad\\
 The number  of $n$-grasping supports with $K$ tetrahedra in a triangulated manifold with $N$ tetrahedra is given by the number of cardinality $K$ subsets of the set of tetrahedra of $M$, in other words by $\frac{N!}{(N-K)!K!}$.  
On the other hand:
\begin{multline}\label{B}
\frac{1}{N^n}{\V^{(n)}(}M,\D)\\= \frac{1}{{4^n}N^n}\sum_{K=1}^n \quad \sum_{\substack{
    n-\textrm{grasping supports }  \Tc\\ \textrm{with } K \textrm{
      tetrahedra}}}\quad\sum_{\substack{n-\textrm{graspings } \Gc \textrm{ of
    } M \\ \textrm{supported in } \Tc}}X(M,\D,K,\Gc),
 \end{multline}
where, according to equation (\ref{xm}), 
\begin{multline*}
X(M,\D,K,\Gc)\\=
\sum_{\textrm{colourings } c \textrm{ of } \Tc^1_{\Gc} } {A(\Gc,c)} \dim_q(c){\n}^{-\frac{n_3'}{2} -\frac{\# \Tc^1_\Gc}{2}} Z_\WRT\left(M \#_{{\rho(}\Tc^1_{\Gc})} {\overline{M}},R_\Gc\right).
\end{multline*}

\noindent{\bf Assumption 1} {\it Fix a $3$-manifold $M$ and a  positive integer $n$. Suppose that there exists a positive constant $C=C(M,n)< +\infty$ for which we have that $|X(M,\D,K,\Gc)|\leq C$, for any triangulation $\D$ of $M$, any $K\in \{1,\ldots n\}$ and any $n$-grasping $\Gc$ of $M$ with $K$ tetrahedra.}

\quad\\
\noindent The number of $n$-graspings that can supported in  an arbitrary set with $K$ tetrahedra, with $1\leq K\leq n$, is certainly bounded by a positive constant $C'<\infty$.
As $\D \to 0$, the number of tetrahedra of $M$ goes to infinity. Therefore if $K<n$ then 
\begin{multline*}
\left |\frac{1}{N^n}\quad \sum_{\substack{
    n-\textrm{grasping supports }  \Tc\\ \textrm{with } K \textrm{
      tetrahedra}}}\quad\sum_{\substack{n-\textrm{graspings } \Gc \textrm{ of
    } M \\ \textrm{supported in } \Tc}}X(M,\D,K,\Gc)\right|\\\leq \frac{N!}{{N^n}(N-K)!K!} CC' \to 0 
\end{multline*}
if  $N \to +\infty.$ 

Let us now consider $n$-graspings of $M$ living in $n$ tetrahedra. Such graspings are called separated if the tetrahedra of its support are pairwise disjoint. It is complicated to determine the exact number of separated $n$-graspings with $n$-tetrahedra. This is because this is highly dependent on the local configuration of the chosen triangulation of $M$.

\quad \newline
\noindent {\bf Restriction 2}  {\it Choose a positive integer {$D=D(M)$.} We consider only triangulations $\D$ of $M$  such that  any tetrahedra of $M$ intersects at most ${D}$ tetrahedra of $M$.}

\quad\\
\noindent As we will see below in Subsection \ref{exact}, there exists a positive integer ${D}$ {for which} any $3$-manifold $M$ has a triangulation  such that  any tetrahedra of it intersects at most ${D}$ tetrahedra, and triangulations like this can be chosen to be arbitrarily fine. 

{Restricting to this type of triangulations, the number of separated $n$-grasping supports is  not smaller than $\frac{N(N-{D})(N-2{D})\ldots (N-n{D}+{D})}{n!}$, whereas the number of $n$-grasping supports with $n$-tetrahedra is  $\frac{N(N-1)(N-2)\ldots (N-n+1)}{n!}$.}

Going back to equation (\ref{B}), the value of
\begin{equation*} 
\frac{1}{N^n} \quad \sum_{\substack{
    n-\textrm{grasping supports }  \Tc\\ \textrm{with } n \textrm{
      tetrahedra}}}\quad\sum_{\substack{n-\textrm{graspings } \Gc \textrm{ of
    } M \\ \textrm{supported in } \Tc}}X(M,\D,n,\Gc)
 \end{equation*}
splits into the contribution of separated and non-separated $n$-graspings with $n$ tetrahedra. Since by Assumption 1 the set of possible values of $X(M,\D,n,\Gc)$ is bounded, the contribution of non-separated configurations goes to zero as the number {$N$ of tetrahedra of $M$} goes to $+\infty$. Therefore we have:
\begin{multline}\label{E}
\lim_{N\to \infty}\frac{1}{N^n}{{\V^{(n)}}(}M,\D)\\ = {\lim_{N \to \infty} \frac{1}{{4^n}N^n}  \sum_{\substack{\textrm{separated }
    n-\textrm{grasping supports }  \Tc\\ \textrm{with } n \textrm{
      tetrahedra}}}\quad\sum_{\substack{n-\textrm{graspings } \Gc \textrm{ of
    } M \\ \textrm{supported in } \Tc}}X(M,\D,n,\Gc).}
\end{multline}

Now, the value of $X(M,\D,n,\Gc)$ is {in fact} independent of the chosen separated $n$-grasping $\Gc=\{T_i,\g_i\}_{i=1}^n$ of $M$, supported in the set $\Tc_\Gc=\{T_i\}_{i=1}^n$ of $n$ non-intersecting tetrahedra of $M$; compare with {Conjecture $A$} on page $8$ of \cite{Ba2}. Note that space orderings are not relevant in this case.

Let us see why it is so. By equation  (\ref{xm}) it follows that:
\begin{equation}
X(M,\D,n,\Gc)= 
\sum_{\textrm{colourings } c \textrm{ of } \Tc^1_\Gc }{A(\Gc,c)} \dim_q(c){\n}^{-\frac{7}{2}n} Z_\WRT\left(M \#_{{\rho(}\Tc^1_\Gc)} {\overline{M}},R_\Gc\right),
\end{equation}
since $n_3'$ is the number of vertices of $M$ which are  endpoints of edges in  $\Tc_\Gc^1$, and in this case $\Tc^1_\Gc$ is, topologically, the disjoint union of $n$ $\Y$-graphs, {each with a unique trivalent vertex and three open ends (univalent vertices).}

Given that any two embeddings of a disjoint union of $n$ $\Y$-graphs into $M$ are isotopic, it thus follows that the value of $X(M,\D,n,\Gc)$ is independent of the chosen separated $n$-grasping $\Gc$ with $n$ tetrahedra. Therefore, by the same argument as in the proof of Theorem \ref{volc}, and by using equation (\ref{refer}), it follows that (whenever $\Gc$ is separated)
\begin{multline*}
X(M,\D,n,\Gc)\\={\n}^{-3n}\left(\sum_{a,b,c} \dim_q(a)\dim_q(b)\dim_q(c){{A}_a {A}_b {A}_c} \left <{{{\hat{\Y}}}};a,b,c,1\right>\right)^nZ_\TV(M){=0.}
\end{multline*}
Here ${\hat{\Y}}$ is the graph of Figure \ref{haty}.

There are exactly $4^n$ grasping supported on a set of $n$ non-intersecting tetrahedra. On the other hand, the number of separated $n$-graspings supports is certainly between $\frac{N(N-{D})(N-2{D})\ldots (N-n{D}+{D})}{n!}$ and $\frac{N(N-1)(N-2)\ldots (N-n+1)}{n!}$. Putting everyting together follows that (should Assumption 1 hold true), and restricting to triangulations satisfying Restriction 2 that
$${\frac{1}{N^n}{\V^{(n)}(}M,\D)\longrightarrow\\ {0}.}$$
whenever the number $N$ of tetrahedra of a triangulation $\D$ of $M$ converges to $+\infty$. 
This finishes a preliminary analysis of the Dilute Gas Limit.

\subsubsection{The problem with {Assumption 1} }

Fix a positive integer $n$ and a 3-dimensional manifold $M$.
We would like to prove that there exists a positive constant $C=C(M,n)<\infty$ such that $|X(M,\D,K,\Gc)|\leq C$, for any $n$-grasping $\Gc$ of $M$ with $K$ tetrahedra, where $K=1,2,\ldots n$, and an arbitrary triangulation $\D$ of $M$. This is a difficult problem.

The approach taken in \cite{Ba2} was to conjecture that $X(M,\D,K,\Gc)$ can only take a finite number of values, for fixed $M$ and $n$. However, this is very likely to be false {in our case.} Indeed, as we have seen above, 
\begin{multline*}
X(M,\D,K,\Gc)\\=
\sum_{\textrm{colourings } c \textrm{ of } \Tc^1_{\Gc} }{A(\Gc,c)} \dim_q(c){\n}^{-\frac{n_3'}{2} -\frac{\# \Tc^1_\Gc}{2}} Z_\WRT\left(M \#_{{\rho(}\Tc^1_{\Gc})} {\overline{M}},R_\Gc\right),
\end{multline*}
{where $\Gc=\{(T_i^\Gc,\g_i^\Gc,\O_i^\Gc)\}_{i=1}^K$, $\Tc^1_\Gc$ is defined in equation (\ref{F}) and ${A(\Gc,c)}$ is defined in equation (\ref{GG}).} {For fixed $K$ and $n$, where $K$ is large enough}, considering the set of all triangulations $\D$ of $M$, there can be an infinite number of isotopy classes of sets of $M$ that are of the form $\Tc^1_\Gc$ for some $n$-grasping $\Gc=\{(T_i,\g_i,\O_i)\}_{i=1}^K$  with $K$ tetrahedra. 

{For example, consider a  triangulation of the solid torus with $K$ tetrahedra, with a  grasping in each, so that all edges of these $K$ tetrahedra are incident to some 	grasping. Then embed the solid torus into the manifold $M$ (there {exists an infinite} number of such embeddings) and extend the triangulation of the solid torus to a triangulation of $M$ (this can always be done).}

Therefore  $X(M,\D,K,\Gc)$ can almost certainly  take an infinite number of values for fixed $n$ and $M$, from which we can assert that Conjecture $A$ in page 8 of \cite{Ba2} {is} probably false in our particular case\footnote{The number of topologically distinct possible classes for $(\Tc^1_\Gc,R_\Gc)$ is infinite; however, there still exists the unlikely possibility that $Z_\WRT\left(M \#_{{\rho(}\Tc^1_{\Gc})} {\overline{M}},R_\Gc\right)$ may take only a finite number of values.}. This makes it difficult to give an upper bound for $X(M,\D,K,\Gc)$. 

To fix  this problem we will alter slightly the way of defining the limit (\ref{Lim}), by restricting the class of allowable triangulations.

\subsection{Exact Calculation}\label{exact}

Since they are easier to visualise, we will now switch to cubulations of $3$-manifolds. Fix a closed $3$-manifold $M$. A cubulation of $M$ is a partition of it into $3$-cubes, such that if two cubes intersect they will do it in a common face, edge or vertex of each. Note that any $3$-manifold  can be cubulated.

Any cubulation $\square$ of  $M$ will give rise to a triangulation $\D_\square$ of it by taking the cones first of  each face and then of each  cube of $M$.
 Given a cubulation $\sq$ of $M$ we  therefore define 
$${{\V^{(n)}}(}M,\sq)={{\V^{(n)}}(}M,\D_\sq).$$
Here $n$ is a positive integer.

The three dimensional cube can  be naturally subdivided into {8} cubes. This will be called the baricentric subdivision. The baricentric subdivision of a cubulated manifold is obtained  by doing the baricentric subdivision of each cube of it. Denote the $v^{\rm th}$ baricentric subdivision of $M$ by $\sq^v$. 

 Fix a cubulation $\sq$ of $M$. We want to consider the limit: 
\begin{equation}\label{Lim2}
\lim_{v \to +\infty} \frac{1}{{N}_v^n}{{\V^{(n)}}(}M,\sq^v),
\end{equation}
where ${N}_v$ is the number of tetrahedra of the triangulation $\D_{\sq^v}$ of $M$.

 We want to use the calculation in \ref{pre}. Therefore Assumption 1 and Restriction 2 need to be addressed. Their validity is, as we have seen, highly dependent on the local combinatorics of the chosen triangulations. Therefore, we make the following restriction on the cubulations with which we work. 

\begin{Definition}[Acceptable cubulation]
The valence of an edge in a cubulated  3-manifold $M$ is given by the number of cubes in which the edge is contained. 
A cubulation of the 3-manifold $M$ is called acceptable if any edge of $M$ has valence $3,4$ or $5$, and the set of edges of order $3$ and of order $5$ match up to form 1-dimensional disjoint submanifolds $\S_3$ and $\S_5$ of $M$.
\end{Definition}
It is proved in \cite{CT} that any closed orientable 3-manifold has an acceptable cubulation. Note that if $\sq$ is acceptable then so is the baricentric subdivision of it; see below.

Let us try to visualise an acceptable cubulation of $M$. The manifolds $\S_3$ and $\S_5$ are  disjoint union of circles $S^1$.  Let $\S$ be a component of $\S_3$. The cubical subcomplex $\hat{\S}$ of $M$ made from the cubes  of  $M$ which contain some simplex of $\S$, together with their faces, is diffeomorphic to $D^2\times S^1$. Moreover $\hat{\S}$ is cubulated  as the product of the tri-valent cubulation of the disk $D^2$, shown in {Figure} \ref{model}, with some cubulation of $S^1$. An analogous picture holds if $\S$ is a component of $\S_5$, by using the penta-valent cubulation of the disk $D^2$ shown in {Figure} \ref{model}.
Then what is left of the cubulation of $M$ is locally given by some portion of the natural cubulation of the $3$-space (with vertices at $\Z \times \Z \times \Z$.)

From this picture, it is easy to see that if $\sq$ is an  acceptable cubulation of $M$ then so is the baricentric subdivision of it.  Moreover, we can show that there exists a positive integer ${D}$ such that, for any $3$-manifold with an acceptable cubulation $\sq$, then any tetrahedron of $\D_\sq$ intersects at most ${D}$ tetrahedra of $\D_\sq$. This proves that Restriction 2 will hold if we consider triangulations coming from taking the cone of  acceptable cubulations.
\begin{figure}
\centerline{\relabelbox 
\epsfysize 2.5cm
\epsfbox{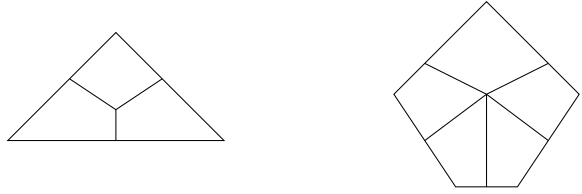}
\endrelabelbox } 
\caption{\label{model} Tri-valent and penta-valent cubulations of the disk $D^2$.}
\end{figure}

Looking at Assumption 1, let us now prove that given an acceptable cubulation $\sq$ of $M$, then 
\begin{multline*}
X(M,\D_{\sq^v},K,\Gc)\\=  \sum_{\textrm{colourings } c \textrm{ of } \Tc^1_{\Gc} }{A(\Gc,c)} \dim_q(c){\n}^{-\frac{n_3'}{2} -\frac{\# \Tc^1_\Gc}{2}} Z_\WRT\left(M \#_{{\rho(}\Tc^1_{\Gc})} {\overline{M}},R_\Gc\right)
\end{multline*}
can only take a finite number of values, for fixed $M$ and $n$. Here $v$ is an arbitrary  positive integer, and $\Gc=\{(T_i^\Gc,\g_i^\Gc,\O_i^\Gc)\}_{i=1}^K$ is an $n$-graping of $M$ with $K$ tetrahedra, thus $K\in \{1,\ldots, n\}$. Recall also that $\Tc^1_{\Gc}$ is given by equation (\ref{F}) and $n_3'$ denotes the number of vertices of the graph made out of the edges of $\Tc^1_\Gc$, together with their endpoints. In particular ${\n}^{-\frac{n_3'}{2} -\frac{\# \Tc^1_\Gc}{2}}$ can only take a finite number of values.

 On the other hand,  the term  
$$\sum_{\textrm{colourings } c \textrm{ of } \Tc^1_{\Gc} }{A(\Gc,c)} \dim_q(c)  Z_\WRT\left(M \#_{{\rho(}\Tc^1_{\Gc})} {\overline{M}},R_\Gc\right)$$
 depends only  on the isotopy class of the pair $(\Tc^1_\Gc,R_\Gc)$ inside $M$. The following lemma shows that there exists only a finite number of possible isotopy classes of  $\Tc^1_\Gc$ in $M$ for a fixed $n$. There exist also a  finite (and depending only on $n$)   
number of possible configurations of the graph $R_\Gc$, wrapping around $\Tc^1_\Gc$. 
\begin{Lemma}\label{C}
Let $M$ be a 3-dimensional manifold with an acceptable cubulation $\sq$. Let $Q$ be a fixed positive integer. There exists a finite number of possible isotopy  classes of graphs in $M$ which can be constructed out	 of $Q$ edges of the triangulation $\D_{\sq^v}$ of $M$, where $v$ is arbitrary.
\end{Lemma}

By using this lemma {(proved in Subsection \ref{D}),} the same argument as in Subsection \ref{pre} shows the following theorem:
\begin{Theorem}\label{Main}
Let $M$ be an oriented closed $3$-manifold. Let $\sq$ be an acceptable cubulation of $M$. Let $n$ be a positive integer. {If $N_v$} denotes the number of tetrahedra of the triangulation $\D_{\sq^v}$, then:
$$ {\lim_{v \to+ \infty} \frac{1} {N_v ^n} {\V^{(n)}(}M,\sq^v)=0\,.}$$

\end{Theorem}

Therefore in the dilute gas limit {with $g=\lambda N$:}
$$ {Z(M)= \sum_{n=0}^\infty (-1)^n \lambda^n \V^{(n)}(M) \to \sum_{n=0}^\infty  \frac{g^n}{n!} z_1^n Z_\TV (M) = e^{g z_1}Z_\TV(M)= Z_\TV (M)\,,}$$
{since} 
$$ {z_1= -{\n}^{-3} \sum_{a,b,c} \dim_q(a)\dim_q(b)\dim_q(c){{A}_a {A}_b {A}_c} \left <{\hat{\Y}};a,b,c,1\right>=0\,,}$$
{see the discussion in the Introduction.}

\subsection{{Dilute gas limit for $z_1 =0$}}

The fact that $z_1 =0$ implies that the dilute gas limit partition function for {$g=\lambda N$} is the same as the unperturbed one. In order to obtain a non-trivial partition function we need to take a different dilute gas limit.

Let us now consider configurations where a single tetrahedron of the manifold $M$ contains a  space ordered 2-grasping $\Gc$. Let $\T\subset S^3$ be the standard tetrahedron. Let $\rho(\T)$ be a regular neighbourhood of $\T$ in $S^3$. Let also $R_\Gc\subset \d \rho(\T)$ be the associated graph with crossing information in the boundary of $\rho(\T)$; see subsection \ref{Hoc}. 

In the general case when there are $p$ graspings in a single tetrahedron $T$ of $M$ one  defines
\begin{multline}\label{zp} 
z_p =\frac{1}{(-4)^p}\sum_{\substack{\textrm{space ordered }\\ p\textrm{-graspings } \Gc \textrm{ of } \T}}  \sum_{\textrm{colourings } c \textrm{ of } \T }\\{A(\Gc,c)}
\dim_q(c){\n}^{-4} Z_\WRT\left(S^3\#_{{\rho(}\T)} {\overline{S^3}},R_\Gc\right)\,;
\end{multline}
see equation (\ref{x2}). Note there exists a unique way to embeed an oriented tetrahedron in $M$, up to isotopy.  From the proof of theorem \ref{volc}, for $p=1$ this coincides with the previous definition of $z_1=0$. For $p=2$, this reduces to
\begin{multline}\label{ztwo}
 z_2 = \frac{{\n}^{-3}}{4} \sum_{a,b,c} \dim_q(a)\dim_q(b)\dim_q(c) {A}_a {A}_b {A}_c \langle {Y_2};a,b,c,1\rangle\\
+\frac{3{\n}^{-3}}{8}\sum_{a_1,a_2,a_3,a_4,e}\quad \left(\prod_{i=1}^4 {A}_{a_i} \dim_q(a_i)\right) {A}_e^2 \dim_q(e)  \langle {H_2} ;a,b,c,d,e,1 \rangle \,,
\end{multline}
where ${Y_2}$ is the linear combination of graphs shown in figure \ref{Y2}, while ${H_2}$ is the linear combination  of graphs  appearing in figure \ref{H2}.  The constants $A_i$ {are given by equation (\ref{aj})}. The $Y_2$ graphs correspond to the case when both graspings are associated with the  same tetrahedron vertex, while the $H_2$ graphs correspond to the case when the graspings connect  edges incident to two different tetrahedron vertices.

\begin{figure}
\centerline{\relabelbox 
\epsfysize 3cm
\epsfbox{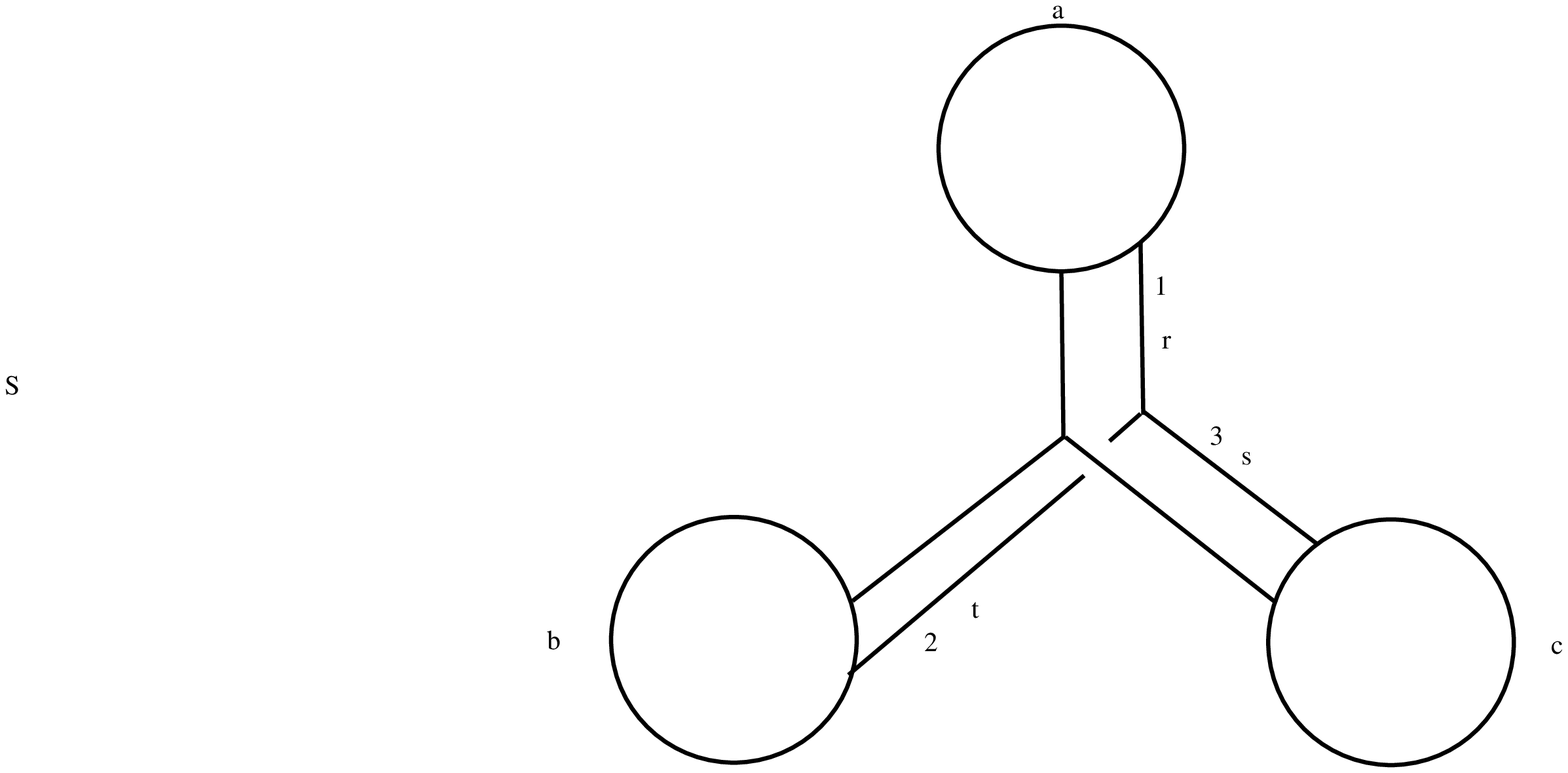}
\relabel{S}{$\displaystyle{\sum_{r,t,s=0}^1}$}
\relabel{r}{$\s{r}$}
\relabel{t}{$\s{t}$}
\relabel{s}{$\s{s}$}
\relabel{a}{$\s{a}$}
\relabel{b}{$\s{b}$}
\relabel{c}{$\s{c}$}
\endrelabelbox}\caption{\label{Y2} The linear combination of graphs ${Y_2}$. The two $\Y$-graphs in the middle are coloured by the spin 1. The indices labeling the bottom $\Y$-graph refer to framing coefficients.}
\end{figure}
\begin{figure}
\centerline{\relabelbox 
\epsfysize 5cm
\epsfbox{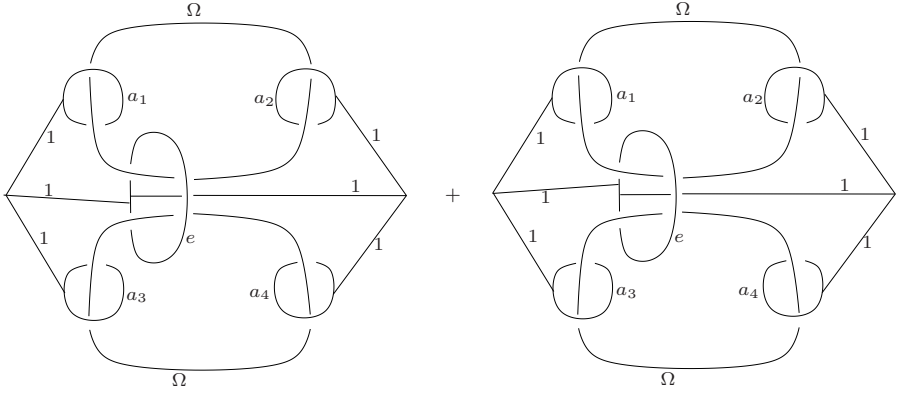}
\relabel{a}{$\s{a_1}$}
\relabel{b}{$\s{a_2}$}
\relabel{c}{$\s{a_3}$}
\relabel{d}{$\s{a_4}$}
\relabel{e}{$\s{e}$}
\relabel{W}{$\s{\W}$}
\relabel{WW}{$\s{\W}$}
\relabel{g}{$\s{1}$}
\relabel{h}{$\s{1}$}
\relabel{i}{$\s{1}$}
\relabel{j}{$\s{1}$}
\relabel{k}{$\s{1}$}
\relabel{l}{$\s{1}$}
\relabel{a1}{$\s{a_1}$}
\relabel{b1}{$\s{a_2}$}
\relabel{c1}{$\s{a_3}$}
\relabel{d1}{$\s{a_4}$}
\relabel{e1}{$\s{e}$}
\relabel{W1}{$\s{\W}$}
\relabel{WW1}{$\s{\W}$}
\relabel{g1}{$\s{1}$}
\relabel{h1}{$\s{1}$}
\relabel{i1}{$\s{1}$}
\relabel{j1}{$\s{1}$}
\relabel{k1}{$\s{1}$}
\relabel{l1}{$\s{1}$}
\relabel{+}{$\s{+}$}
\endrelabelbox}\caption{\label{H2} {The combination of graphs $H_2$.} }
\end{figure}

Let us  consider an even order perturbative contribution
$$  Z_{2n} = {1\over (2n)!}\left(\sum_{m=1}^N \hat V_m \right)^{2n} Z(J)\Big{|}_{J=0} \,,$$
where $\hat V_m = \partial^3_J (\tau_m)$. We write it as
$$ Z_{2n} = {1\over (2n)!}\Big{\langle} \left(\sum_{m=1}^N  V_m \right)^{2n}\Big{\rangle} \,.$$
Then
$$ Z_{2n} = {1\over (2n)!}\sum_{p=1}^{2n}\sum_{1\le m_1, \cdots, m_p\le N} \sum_{k_1,\cdots,k_p}{(2n)!\over k_1 ! \cdots k_p !} \Big{\langle} V_{m_1}^{k_1} \cdots V_{m_n}^{k_p} \Big{\rangle}\,. $$

Let $N$ be the number of tetrahedra of $M$. When $N\to+\infty$ (in the sense described in subsection \ref{exact}), the dominant contribution comes from the graspings supported by $n$ non-intersecting tetrahedrons. This contribution {arises} from the $p=n$ terms in $Z_{2n}$ with $k_1 = k_2 = \cdots = k_n = 2$. By using the same technique as in subsections \ref{pre} and \ref{exact} we can show that
$$\sum_{1\le m_1, \cdots, m_n\le N} {(2n)!\over 2^n}\langle V_{m_1}^2 V_{m_2}^2 \cdots V_{m_n}^2 \rangle \approx {(2n)!\over 2^n} C^N_n z_2^n Z_0\approx {(2n)!\over 2^n}{N^n \over n!} z_2^n Z_0\,,$$
where $Z_0 = Z_{\TV}(M)$. Therefore
$$ Z_{2n} = {1\over 2^n}{N^n \over n!} z_2^n Z_0 + O(N^{n-1})\,,$$
as $N\to\infty$.

Similarly
$$ Z_{2n+1} = -{1\over (2n+1)!}\sum_{p=1}^{2n+1}\sum_{1\le m_1, \cdots, m_p\le N} \sum_{k_1,\cdots,k_p}{(2n+1)!\over k_1 ! \cdots k_p !} \Big{\langle} V_{m_1}^{k_1} \cdots V_{m_n}^{k_p} \Big{\rangle}\,, $$
and the dominant contribution for $N$ large comes from the graspings supported by $n$ non-intersecting tetrahedrons. This contribution corresponds to $p=n$ terms with $k_1 = \cdots = k_{n-1}=2,k_n=3$, and it is easy to show that
$$\sum_{1\le m_1, \cdots, m_n\le N} {(2n+1)!\over 2^{n-1}3!}\langle V_{m_1}^2  \cdots V_{m_{n-1}}^2 V_{m_n}^3\rangle \approx {(2n+1)!\over 2^{n-1}3!} nC^N_n  z_2^{n-1} z_3 Z_0$$
$$\approx {(2n+1)!\over 2^{n-1}3!}{N^n \over (n-1)!} z_2^{n-1} z_3  Z_0\,,$$
where $z_3$ is defined in equation (\ref{zp}).
Therefore
$$ Z_{2n+1} = -{1\over 3\cdot 2^n}{N^n \over (n-1)!} z_2^{n-1} z_3 Z_0 + O(N^{n-1})\,,$$
as $N\to\infty$.

Let ${g=\lambda^2 N}$, and $\bar Z_N = \sum_{n=0}^N \lambda^n Z_n (M,\Delta)$, then
$$\bar Z_N \approx \sum_{n=1}^{N/2} {g^n \over n!} z_2^n Z_0 - {\lambda\over 3}\sum_{n=1}^{N/2} {g^n \over (n-1)!} z_2^{n-1}  z_3 Z_0$$
$$ \approx e^{g z_2}Z_0 - {\lambda g\over 3} z_3 e^{g z_2}Z_0$$
for $N$ large. By taking the limit $N\to\infty$, $\lambda\to 0$ such that {$g=const$,} we obtain
\begin{equation}\lim_{N\to\infty}\bar Z_N (\Delta ,M)= e^{g z_2}Z_0 (M) \,. \label{ppf}\end{equation}

{One can conjecture that in the case $z_1 = z_2 = \cdots = z_{p-1} =0$, $z_p \ne 0$, the limit $N\to 0$, $\l\to 0$, such that $\lambda^p N$ is a non-zero constant, gives
\begin{equation}\lim_{N\to\infty}\bar Z_N (\Delta, M) = e^{g z_p} Z_0 (M)\,, \label{nppf}\end{equation}
where $g\propto \l^p N$.}

\subsection{Relating $g$ and $\l$}

{Note that the perturbed partition function (\ref{nppf}) is a function of the parameter $g$, which takes values independently from the values of $\l$. One would like to find a relation between $\lambda$ and $g$ such that}
$$ Z(M,\l) = e^{gz_p (r)}{Z_{\TV}}(M,r) \,,$$
{where we have written explicitely the dependence of $z_p$ and ${Z_{\TV}}$ on the integer $r$.
If we assume that the path integral (\ref{3dpi}) for $\lambda =(2\pi/r)^2$ is equal to ${Z_{\TV}}(M,r)$, it is natural to consider $\bar\lambda =\lambda - (2\pi/r)^2$ as the perturbative parameter. Let $g = f(\bar\lambda)$, then 
for $\l =(2\pi/l)^2$, where $l$ is an integer different from $r$, we will require that} 
\begin{equation} 
Z \left(M,(2\pi/l)^2\right) =e^{f(\bar\l)z_p (r)}{Z_{\TV}}(M,r) = {Z_{\TV}}(M,l) \,.\label{rone}
\end{equation}
{The relation (\ref{rone}) implies that}
\begin{equation} 
f\left((2\pi/l)^2 - (2\pi/r)^2\right)={1\over z_p (r)}\ln {{Z_{\TV}}(M,l)\over {Z_{\TV}}(M,r)}\,.\label{rtwo}
\end{equation}

{When $2\pi/\sqrt{\l}$ is not an integer, a natural generalization of the relations (\ref{rone}) and (\ref{rtwo}) is}
\begin{equation}
Z(M,\l) = {Z_{\TV}}\left(M,{2\pi\over\sqrt\l}\right)\label{zlamb}
\end{equation}
{and}
\begin{equation} f\left(\l - (2\pi/r)^2\right)={1\over z_p (r)}\ln {{Z_{\TV}}\left(M,{2\pi\over\sqrt\l}\right)\over {Z_{\TV}}(M,r)}\,.\label{flamb}\end{equation}
{Although $f$ depends on $z_p$ and the integer $r$, the value for $Z(M,\l)$ given by (\ref{zlamb}) is independent of $z_p$ and $r$.}

\subsection{Proof of Lemma \ref{C}}\label{D}

{Consider the cubulations $C_3$, $C_4$ and $C_5$  of $\R^2$ presented in {Figures} \ref{exp}. These have the property that they are invariant under baricentric subdivision; see {Figure} \ref{exp2}.} Doing the product with the cubulation of $\R$ with a vertex at each integer, yields cubulations $C_3'$, $C_4'$ and $C_5'$ of $\R^3$, which stay stable under baricentric subdivision. These cubulations of $\R^3$ have the property that, given a positive integer $Q$, then there exists a finite number of isotopy  classes of graphs in $\R^3$ which can be constructed out of $Q$ edges of the triangulations of $\R^3$ constructed  by taking the cone of them. 

Let $M$ be a 3-dimensional manifold with an acceptable cubulation $\sq$. We can cover $M$ with a finite number of cubical subcomplexes, say $\{V_i\}$, each of which is  isomorphic to a subcomplex of either $C_3'$, $C_4'$ or $C_5'$. Moreover, we can choose each $V_i$ {so that it is diffeomorphic to the 3-ball $D^3$.} Suppose that $\G$ is a graph (which we can suppose to be connected) made from $Q$ edges of the triangulation $\D_{\sq^v}$ of $M$, where $v$ is arbitrary. By making $v$ big enough, we can suppose that any such graph $\G$ is contained in $V_i$ for some $i$. This means that $\G$ is isomorphic to a graph with $Q$ edges either in $C_3',C_4'$ or $C_5'$, and there are only a finite number of isotopy classes of these. 

\begin{figure}
\centerline{\relabelbox 
\epsfysize 3.3cm
\epsfbox{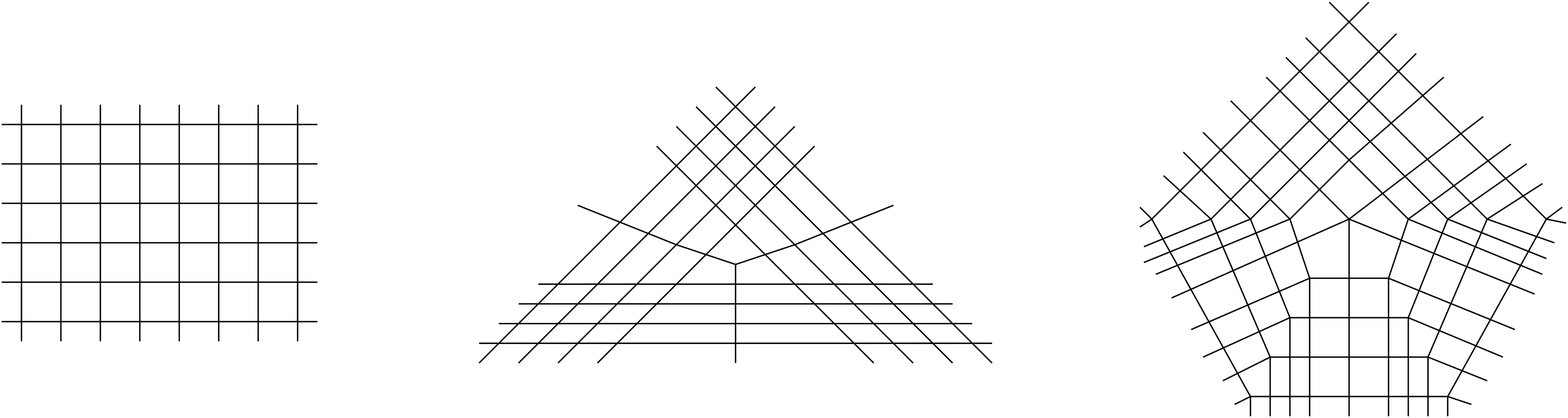}
\endrelabelbox } 
\caption{\label{exp} {The cubulations  $C_4$, $C_3$ and $C_5$ of $\R^2$.}}
\end{figure}
\begin{figure}
\centerline{\relabelbox 
\epsfysize 3.5cm
\epsfbox{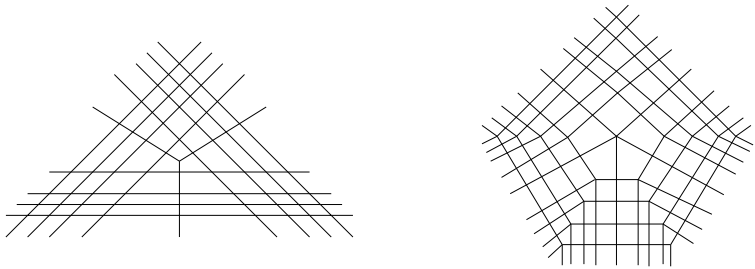}
\endrelabelbox } 
\caption{\label{exp2} {Baricentricaly subdividing   the cubulations $C_3$ and $C_5$ of $\R^2$ yields  $C_3$ and $C_5$.}}
\end{figure}

\section{Conclusions}

Note that 
$${V^{(n)}_\TV (M,\Delta) = \frac{1}{4^n}\sum_{K=1}^n \sum_{\cal G}n(K,{\cal G})\,X(M,\Delta,K,{\cal G} )\,,}$$ 
where $n(K,\cal G)$ is the number of times the configuration $\cal G$ appears, up to isotopy, among configurations which correspond to $n$-graspings distributed among $K$ tetrahedrons. Then
$$V^{(n)}_\TV (M,\Delta)= \sum_{k=0}^n N^{n-k} v_{k+1}^{(n)}(M,\Delta)\,,$$
where the coefficients $v_k^{(n)}$ are linear combinations of {$X(M,\Delta,K,\Gc)$.} In general $v_k^{(n)}$ {depends on the triangulation $\D$,} except {for $v_1^{(n)} = \frac{1}{n!} z_1^n Z_\TV $.} For the triangulations coming from {the baricentric divisions of an acceptable} cubulation, there is only a finite number of topologically distinct grasping configurations. Therefore the set of values of the $X$'s is finite and hence limited, so that {the $v_k^{(n)}$ are} limited as $N\to\infty$. {In that case the dilute gas configurations which have the dominant contributions have two graspings in a single tetrahedron instead of one, because $z_1 =0$ and  $z_2 \ne 0$. The corresponding difference with respect to the Baez definition of the dilute gas limit is that the effective perturbation parameter $g$ changes from {$\l N$ to $\l^2 N$.}}

{We have not proved that $z_2$, which is given by (\ref{ztwo}), is different from zero. However, it is very unlikely that $z_2$ vanishes, since the evaluations of $Y_2$ and $H_2$ graphs are not apparently zero.}

{The result $z_1 =0$ is a consequence of our choice of the volume operator $\hat V$. This is a natural choice since it contains only the non-coplanar triples of the tetrahedron edges. One can also include the coplanar triples, see \cite{HS}, and it is possible that in that case $z_1\ne 0$. This would then give $Z=e^{gz_1}Z_0$ in the usual dilute gas limit.}

{Note that the proposed value for $Z(M,\l)$ when $2\pi/\sqrt{\l}$ is not an integer, given by (\ref{zlamb}), is independent of $z_p$. This means that it is independent of the type of dilute gas limit used. Since the value of $z_p$ dependes on the choice of the volume operator, this also means that the value (\ref{zlamb}) is independent of the choice of the volume operator. Although the value (\ref{zlamb}) is independent of the triangulation of $M$, it} is not a new topological invariant, because it is the same function as ${Z_{\TV}}(M,r)$. The only diference is that $r=2\pi/\sqrt{\l}$ takes a noninteger value. However, (\ref{zlamb}) gives a definition of 3d quantum gravity partition function when the value of the cosmological constant is an arbitrary positive number.

An interesting problem is to develop the PR model perturbation theory without using the quantum group regularization.
The recent results on the PR model regularisation by using  group integrals, see \cite{BNG}, suggest that such a perturbation theory could be developed. The corresponding perturbative series could be then summed by using the dilute gas techniques and the cubulation approach. The obtained result could be then compared to $Z_\TV (M,r)$.

The techniques developed in this paper can be readily extended to the case of four-dimensional Euclidean Quantum Gravity with a cosmological constant, since then the classical action can be represented as the ${\rm SO}(5)$ BF theory action plus a perturbation quadratic in the $B$ field, see \cite{M1,M2}. 

\section*{Acknowledgements}
{We would like to thank Laurent Freidel for helpful comments.}
{J. Faria Martins was supported by the Centro de Matem\'{a}tica da
Universidade do Porto {\it www.fc.up.pt/cmup}, financed by FCT through the programmes POCTI
and POSI, with Portuguese and European Community structural funds, and by
the research project POCTI/MAT/60352/2004, also}
financed by the FCT. 
A. Mikovi\'{c} was partially supported by the FCT grant PTDC/MAT/69635/2006.

\end{document}